\documentclass[a4paper,fleqn,usenatbib]{mnras}
\usepackage[T2A, T1]{fontenc}
\usepackage{ae,aecompl}
\usepackage{graphicx}
\usepackage{amsmath}
\usepackage{amssymb}
\usepackage{array,tabularx}
\usepackage{ulem}
\usepackage[english]{babel}
\usepackage{threeparttable}
\usepackage{enumitem}
\makeatletter
\let\TPT@hookin\@gobble
\let\TPT@hookarg\@gobble
\makeatother

\title[Sco~X-1: parameters]{Parameters of X-ray binary system Scorpius X-1}

\author[A. M. Cherepashchuk et al.]{A. M. Cherepashchuk\thanks{E-mail:Cherepashchuk@gmail.com (AMC)},
T. S. Khruzina\thanks{E-mail:kts@sai.msu.ru (TSK)},
and A. I. Bogomazov\thanks{E-mail:a78b@yandex.ru (AIB)}
\\
M. V. Lomonosov Moscow State University, P. K. Sternberg Astronomical Institute, 119234, Universitetkij prospect, 13, Moscow, Russia\\
}

\date{Accepted . Received ; in original form }

\pubyear{2021}

\begin{document}
\label{firstpage}
\pagerange{\pageref{firstpage}--\pageref{lastpage}}
\maketitle

\begin{abstract}
We modelled optical light curves of Sco~X-1 obtained by the Kepler space telescope during K2 mission. Modelling was performed for the case of the strong heating of the optical star and accretion disc by X-rays. In the considered model the optical star fully filled its Roche lobe. We investigated the inverse problem in wide ranges of values of model parameters and estimated following parameters of Sco X-1: the mass ratio of components $q=M_x/M_v=3.6$ ($3.5-3.8$), where $M_x$ and $M_v$ were masses of the neutron and optical stars correspondingly, the orbital inclination was $i=30^{\circ}$ ($25^{\circ}-34^{\circ}$). In the brackets uncertainties of parameters $q$ and $i$ were shown, they originated due to uncertainties of characteristics of the physical model of Sco X-1. The temperature of non-heated optical star was $T_2 = 2500-3050$ K, its radius was $R_2=1.25R_{\odot}=8.7\times 10^{10}$ cm, and its bolometric luminosity was $L_{bol}=(2.1-4.6)\times 10^{32}$ erg s$^{-1}$. The mass of the star was $M_v\simeq 0.4M_{\odot}$. The contribution of the X-ray heated accretion disc dominated in the total optical emission of Sco~X-1. The transition between low and high states occurred due to the increase of X-ray luminosity by a factor $2-3$.
\end{abstract}

\begin{keywords}
binaries: close -- stars: neutron -- stars: individual: Sco~X-1 -- accretion -- accretion discs
\end{keywords}

\section{Inroduction}

A persistent low mass X-ray binary system Sco~X-1 = V818~Sco was the first compact X-ray source found outside the Solar system \citep{giacconi1962}. A model of an X-ray binary with a neutron star (NS) for Sco~X-1 was suggested by \citet{shklovskii1968}, but the final approval of this model was made later \citep{gottlieb1975,cowley1975}. The cause of the delay was a high irregular variablity of the object V818 Sco (B=$11.1^m-14.1^m$, see, e. g., \citealp{sandage1966,hiltner1967,hiltner1970,bradt1975,canizares1975,mook1975}), and it was very difficult to find any periodicity in brightness changes and in spectra of this source. Numerous studies of Sco~X-1 in X-ray, optical and radio ranges (see, e. g., a catalogue by \citealp{cherepashchuk1996}) made it possible to understand main features of this system of the Z-sources subclass \citep{hasinger1989}, the subclass of bright X-ray sources of the bulge. In the X-ray diagram ``colour in the soft range --- colour in the hard range'' \citep{hasinger1989} the corresponding locus was similar to the letter ``Z'' with three branches: a horizontal branch (HB), a normal branch (NB), and a flaring branch (FB).

As a typical Z-source Sco~X-1 showed the X-ray flux in the lower part of the Z-diagram close the Eddington limit, this fact was used to estimate the distance to Sco~$\textrm{X-1}$ as $d=2\pm 0.5$ kpc, and the corresponding colour excess $E(B-V)\simeq 0.30^m$ (see, e. g., \citealp{cherepashchuk1996}). Sco~X-1 showed quasi-periodical oscillations of the X-ray radiation with frequencies  about 6.3 Hz in NB, about 14.4 Hz in the lower part of FB, and about 10-20 Hz in HB. Herewith the accretion rate onto the NS monotonically grew along the Z-shape curve in the diagram of X-ray colours from $0.4\times 10^{-8}M_{\odot}$ yr$^{-1}$ in HB and $0.6\times 10^{-8}M_{\odot}$ yr$^{-1}$ in NB up to $1.1\times 10^{-8}M_{\odot}$ yr$^{-1}$ in FB \citep{vrtilek1991}. There were also alternative interpretations of Z-digrams, where the accretion rate did not increase monotonically (see, e.g., \citealp{church2012}). It was generally accepted that Z-sources have higher accretion rate in comparison with numerous atoll (``island'') sources with sub-Eddington accretion.

The optical variability of Sco~X-1 showed bimodal and even trimodal character (see, e. g., \citealp{bradt1975,canizares1975,mook1975}). Around the lowest optical flux the optical variability was anti-correlated with the X-ray flux, during the brightening of Sco~X-1 there was a correlation of the optical flux with the X-ray flux variability within FB branch of Z-diagram (see, e. g., \citealp{ilovaisky1980,petro1981,augusteijn1992,mcnamara2003,mcgowan2003}). The analysis of the correlation between X-ray and optical variabilities of Sco~X-1 and estimates of characteristic delay times of the optical variability in comparison to the X-ray variability ($\Delta t\simeq 10\pm 5$ s) was performed by \citet{m-d-2007,britt2013}. The correlation between optical and X-ray variabilities was observed only if Sco~X-1 was in the bright X-ray state and was located in FB branch of the X-ray Z-diagram. A regular optical variability of Sco~X-1 with a period $\approx 18.9$ h was discovered using archival photographic plates by \citet{gottlieb1975} and was confirmed using spectroscopic observations \citep{cowley1975}.

\citet{scaringi2015,hakala2015,hynes2016} conducted the analysis of optical observations of Sco~X-1 obtained by Kepler K2 mission and of X-ray observations obtained by Fermi GBM and MAXI. An average wavelength of wide wavelength range observations of Kepler K2 mission was in the middle of a band that corresponded to {\it B}, {\it V}, {\it R} filters. Optical observations of Sco~X-1 by Kepler in its K2 program were performed during 78.8~d in August-November 2014. An exposure time of an individual observation was 54.2 s. \citet{hynes2016} made careful analysis of random and systematic errors of Kepler K2 observations and extracted 115 680 individual measurements for Sco~X-1. Those data allowed to construct a light curve of this object with 1\% precision. Using the standard sinusoidal light curve \citet{hynes2016} independently found the photometric period of Sco~X-1 as $P=0.78747\pm 0.00072$ d, which was in a good agreement (within the errors) with the more exact spectroscopic period $P=0.7873114\pm 0.0000005$ d measured using the Doppler shift of narrow emission Bowen lines NIII/CIII \citep{galloway2014}.

Folding of all optical Kepler K2 observations on the spectroscopic period showed that the average modulation of Sco~X-1 brightness was the one wave during the one orbital period (``reflection effect'', see \citealp{cherepashchuk1972,lyutyi1973}) and could be clearly splitted in two states: high and low \citep{hynes2016}. The amplitude of the regular orbital optical light curve obtained by Kepler K2 in high and low states (in the supposition of the sinusoidal variability) were practically the same $\approx 0.15^m$. The difference between average values of Sco~X-1 brightness in high and low states (in the same supposition) reached up to $\approx 0.4^m$.

After the removal of the orbital trend from individual observations \citet{hynes2016} concluded that in the low state the system showed mostly bright optical flashes with duration $\Delta t\leq 1$ d, whereas in the high state there were fast ($\Delta t =8-16$ min) flashes and slow ($\Delta t=2-5$ d) dips. \citet{hynes2016} presented average optical light curves of Sco~X-1 in high and low states (and an average overall light curve obtained by K2 mission) and qualitatively analyzed them using sinusoidal fits of observed light curves.

It was very interesting to make detailed models of those very valuable observational data about Sco~X-1 using modern mathematical descriptions of interacting binary systems \citep{khruzina2001,khruzina2003a,khruzina2003b,khruzina2005,khruzina2015,bisikalo2005,lukin2017,cherepashchuk2019a,cherepashchuk2019}.

\section{Optical light curves (orbital)}

\citet{hynes2016} obtained average light curves of Sco~X-1 in white light using Kepler K2 data, also they used spectroscopic data by \citet{galloway2014}: the epoch of the lower conjunction of the optical star $T_0=\textrm{HJD}\, 245 4635.3683\pm 0.0012$, the spectroscopic orbital period $P=0.7873114\pm 0.0000005$ d, the semi-amplitude of the radial velocity curve of the optical star (obtained using Doppler shifts of narrow Bowen emission lines NIII/CIII) $K=74.9\pm 0.5$ km s$^{-1}$.

\citet{hynes2016} conducted a qualitative analysis of the shape and amplitude of orbital light curves in low and high states and in average. It was noted that all three curves were quasi-sinusoidal, there were no significant differences between amplitudes in high and low states (the full amplitude was $0.147^m \pm 0.012^m$ in the low state and $0.151^m \pm 0.008^m$ in the high state). It was possible to suspect a small phase shift to lower phases for the low state ($\Delta \varphi= 0.032\pm 0.024$ in units of the orbital period), but this shift was in $2\sigma$ level, so, it seemed insignificant. \citet{hynes2016} also made careful analysis of the influence of random and systematic deviations on the shape of orbital light curves.

\begin{figure}
\center
\includegraphics[width=\columnwidth]{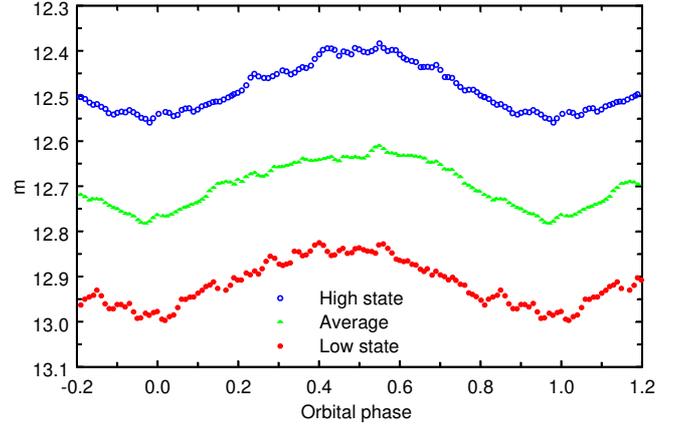}
\vspace{5pt} \caption{Light curves of Sco~X-1 in low and high states (and an average curve) in stellar magnitudes calculated using Equation \ref{eq0}.}\label{c-f-2}
\end{figure}

For further analysis we digitized data from Figure 3 by \citet{hynes2016} using a special computer program, the result was attached in a form of an electronic table. \citet{hynes2016} presented optical light curves of Sco~X-1 as the linear dependence of the quantity of counts of photo detector $N(t)$ on time $t$. To convert the number of counts to stellar magnitudes we used the average quantity of counts (130~000 counts s$^{-1}$), that was close to the average state of the system \citep{hynes2016}. Using photometric data by \citet{galloway2014} obtained in 2001-2009 we were able to attribute the stellar magnitude $V=12.7^m$ to the average value 130~000 counts s$^{-1}$. For the period of K2 observations AAVSO catalogue\footnote{https://www.aavso.org/} contained data about Sco X-1 between 1 August and 11 October 2014 (that partially covered Kepler's observations). The optical brightness $V$ was in the range $12.0^m-13.233^m$, the average value between the maximum and minimum of these values was $12.62^m$, and the average value of all available 58 points in this period of time was $12.34^m$. Both quantities were close to our assumption ($V=12.7^m$). Moreover, since for our calculations relative changes in the light curves were important the result weakly depended on the absolute calibration of light curves. After the re-calculation of the linear count scale to the stellar magnitude scale using formula

\begin{equation}
m=12.7-2.5\log\left(\frac{N}{130\,000}\right),
\label{eq0}
\end{equation}
 
\noindent we obtained orbital light curves of Sco~X-1 in stellar magnitudes. They can be found in Figure \ref{c-f-2}.

\section{Mathematical model of the system}

To interpret light curves we used a model of an interacting binary system that took into account results of three dimensional gas-dynamic calculations \citep{bisikalo2005,lukin2017}. In this model the region of the interaction of the gas stream with the outer border of the accretion disc was a combination of a hot line oriented along the gas stream and a hot spot within the outer border of the accretion disc. A detailed description of this model was given by \citet{khruzina2011}, applications of the model to cataclysmic and X-ray binaries were made by \citet{khruzina2001,khruzina2003a,khruzina2003b,khruzina2005,khruzina2015,cherepashchuk2019a,cherepashchuk2019}.

In Sco~X-1 there was a strong X-ray heating effect $L_x^{\textrm{max}}/L_{opt}\approx 500$, so the role of gas-dynamic interactions of the stream and disc was insignificant in comparison with it. Therefore, in spite the fact that the model in its general form contained around 16 free parameters (which could be found if the system had eclipses), we used a mathematical model with six free parameters.

In our mathematical model we used the standard method of synthesis of light curves of close binary systems by \citet{wilson1971}. The optical donor star filled its Roche lobe. The model took into account the linear limb darkening and gravitational darkening ($\beta=0.08$, $T\sim g^{\beta}$, $g$ was the free fall acceleration at the star's surface, \citealp{lucy1967}), also it took into account the heating of the surface of the star and accretion disc by the X-ray radiation. Fluxes from elementary areas on the star's surface, on the accretion disc and within the interaction region between the gas stream and the disc were calculated using the Planck's law with corresponding local temperatures.

Due to the strong X-ray heating in the Sco~X-1 system the temperature of the non-disturbed (non-heated) optical star $T_2$ cannot be set {\it a priori}, it should be found in the course of the inverse problem solution. The mass ratio $q=M_x/M_v$ ($M_x$, $M_v$ were the masses of the relativistic object and the optical star correspondingly) also cannot be fixed, because, in particular, the correction of the orbital radial velocity curve (that can take into account the asymmetric position of the place of origin of Bowen lines) strongly depended on the model \citep{m-d-2005,galloway2014,hynes2016}. It should be noted that in case of the strong X-ray heating the task of the optical light curve interpretation became sensitive to changes of $q$, because $q$ depended on the radius of the donor star (i.e., on the part of X-ray flux that fell on the star).

The leading role in interpretations of orbital light curves of Sco~X-1 belonged to the account of the ``reflection effect'' and conversion of the energy of the central source in the accretion disc.

The X-ray heating for the optical star was considered in the assumption of an extended central X-ray source irradiating the star isotropically. The temperature of every elemental area on the heated part of the optical star was calculated as the sum of the bolometric flux from the non-disturbed star and of the bolometric flux from the X-ray source:

\begin{equation}
\sigma T^4 = \sigma T^4_0 + (1-\eta_s)F^{bol}_x,
\label{eq-1}
\end{equation}

\noindent where $T$ was the resulting temperature of the area, $F^{bol}_x$ was the falling bolometric flux, $T_0$ was the temperature of non-disturbed area on the star, $\sigma$ was the Stefan-Boltzmann constant, $\eta_s$ was the albedo of the optical star that did not exceed 0.5 according to \citet{dejong1996}.

To calculate the X-ray heating of the accretion disc we also used the model of extended central source as the sphere with a small radius $R_1$ and with the surface temperature $T_{in}$. Parameters $R_1$ and $T_{in}$ were included in the temperature distribution on the non-disturbed accretion disc:

\begin{equation}
T(r)=T_{in}\left(\frac{R_1}{r}\right)^{\alpha_g},
\label{eq-2}
\end{equation}

\noindent where $\alpha_g\leq 0.75$ \citep{shakura1973}. Parameters $R_1$ and $T_{in}$ were searched task parameters of our model. A quasi-parabolic surface of the accretion disc was heated by the slanting X-rays emitted by this sphere. In our model the heating was calculated as the sum of the bolometric flux from the non-disturbed disc (this flux was determined by the gravitational energy release during the accretion) and from the central sphere. In doing so we used a formula that was an analogue of Equation \ref{eq-1}, where the disc's albedo was taken to be equal to $\eta_d=0$ and $\eta_d= 0.9$. As was shown by \citet{dejong1996}, only a small part of X-ray radiation can be reprocessed to the optical range during the X-ray heating of the disc by the slanting X-ray radiation, so $(1-\eta_d)\simeq 0.1$.

The basis of the model of the small extended central X-ray source for the X-ray heating was found by \citet{dubus1999}, who noted that a point X-ray source should be shielded by the accretion disc's body; in order to provide a significant X-ray heating of the outer part of the disc the central source should be outside the disc's plane or the disc should be distorted. We accept the symmetric non-distorted disc, the X-ray heating was calculated in the model of the small extended X-ray source that can be thicker than the inner part of the accretion disc. Also there was a theoretical consideration in favour of the extended central source model \citep{b-k-1977,mitsuda1984,white1985}: the X-ray radiation during the accretion onto the neutron star underwent a significant Compton scattering in the corona of the accretion disc.

It was essential to know a detailed spectrum of X-ray radiation (irradiating the donor star and the accretion disc) to calculate the ``reflection effect'' in lines \citep{antokhina2005}. The presence of the soft component in the X-ray spectrum of the accreting relativistic object led to the formation of emission lines in profiles of absorption lines of the optical star. In addition, the strong X-ray heating of the optical star can lead to a significant outflow of the matter in the form of induced stellar wind from the heated part of the star \citep{basko1973}. Narrow emission lines (caused by Bowen mechanism) were able to form around the starting point of this wind. Also the wind can be emitted from the highly heated accretion disc.

In our case we had orbital light curves of Sco X-1 in a wide range optical continuum. To calculate the ``reflection effect'' we needed only the bolometric luminosity of the source (that strongly changed with time, $L_x=(0.6-12)\times 10^{37}$ erg s$^{-1}$, \citealp{cherepashchuk1996}) as a first approximation. Therefore we limited ourselves by a simple fit of the X-ray source using Planck law, and the average temperature that can be estimated using the observed X-ray luminosity $L_x$. An approximate calculation of the temperature should be performed using the formula $L_x=4\pi\sigma T^4 R^2_{NS}$, where $R_{NS}\approx 10-20$ km was the neutron star's radius. The value of $T$ in this case was $(3-5)\times 10^7$ K. But the size of elementary areas on the surface of the central sphere in our mathematical model of Sco~X-1 was approximately one thousand times greater than the size of the neutron star, so we were forced to average the temperature over surfaces with sizes $\sim 10^3 R_{NS}$, and the bolometric X-ray luminosity $L_x$ remained unchanged in this operation. Therefore the average temperature within the sphere's surface dropped to $T_{in}\sim 10^6$ K, and the model spectrum of X-ray radiation became significantly softer in comparison with the observed spectrum. Nevertheless, calculations of the ``reflection effect'' in continuum required only the bolometric X-ray luminosity of the source, so the described approximation was taken as satisfactory. It allowed to easily parametrize the effectiveness of the X-ray heating of the optical star of the accretion disc using several parameters. They were $T_{in}$ and $R_1$ in Equation \ref{eq-2}, others were the disc's opening angle $\beta_d$ at its outer border (that characterized the size of the shadow from the X-ray source on the surface of the optical star) and the disc's albedo $\eta_d$.

We modelled three optical light curves of Sco X-1: for the low state, for the high state, and the average curve (see Figure \ref{c-f-2}). Transitions between low and high states can be associated either with the change of the intrinsic luminosity of the X-ray source (that in our case followed from the change of the temperature $T_{in}$) or with the shielding of the X-ray radiation of the central source by gas stream structures situated close to the accretion disc or with the change of the irradiation efficiency due to the variable shape of the distorted accretion disc \citep{dubus1999}. Transitions between states we explained by the change of the temperature of the central sphere $T_{in}$.

The value of $T_{in}$ in the low state of Sco X-1 was noted as $T_1$. To solve the inverse problem we started with the grid of $T_1$ values and from the solution of the interpretation of the light curve in the low state we found corresponding values of $R_1$ along with other model parameters. The grid of bolometric luminosity values $L_{bol}^i=4\pi R^2\sigma (T_1^i)^4$ was calculated, then the value of $T_1$ (that corresponded to the X-ray luminosity in the low state) was found.

The X-ray luminosity in the Sco X-1 system greatly changed from one epoch to another, therefore we interpreted orbital optical light curves for the wide range of values of the parameter $T_1=10^5-10^7$ K.

If $T_1$ was $>10^6$ K the bolometric luminosity of the central part of the accretion disc became $L^c_{bol}>10^{38}$ erg s$^{-1}$ and it was greater than the Eddington limit for the neutron star. Nevertheless, due to methodological reasons we used values greater than $10^{38}$ erg s$^{-1}$ to trace the changes of parameters in our inverse problem when the X-ray heating was changed in the wide range.

In general case for the given value of $T_1$ in our inverse problem (as already mentioned above) there were 16 free parameters. But, due to the strong X-ray heating in Sco X-1 it was possible to limit the quantity of free parameters to six: $q$, $i$, $T_2$, $T_{in}$, $R_1/a_0$, $R_d/a_0$, where $T_{in}$ and $R_1$ were parameters of the central sphere (see Equation \ref{eq-2}), $a_0$ was the radius of the system's relative orbit, $R_d/a_0$ was the radius of the accretion disc in units of the radius of the orbit, $T_2$ was the temperature of the non-disturbed optical star.

To solve the inverse problem the Nelder-Mead method was used, we minimized the functional $\Delta$ of the weighted sum of squares of residuals between observed and theoretical curves \citep{himmelblau1972}.

The solution of the inverse problem was performed by variations of one essential parameter. The value of $q$ was fixed, than the minimization of the functional of residuals was made; than the value of $q$ was changed again and the minimization procedure was repeated. As the output of this process the dependence of the minimum of residuals $\Delta_{\textrm{min}}$ on the parameter $q$ was calculated. Using this minimum of residuals the optimal value of $q$ was found. The same procedure was conducted to find the optimal value of the orbital inclination $i$. To avoid getting to a local minimum of the functional of residuals we took a lot (several tens) of initial values for free parameters.

As was pointed out earlier, to interpret the light curve in the low state $T_1$ was accepted to be equal to $T_{in}$, and the minimization of the functional of residuals was performed over five remaining parameters. Finally optimal values of $q$, $i$, $R_1$, $T_2$, $R_d$ for the low state were found, they corresponded to $T_1=T_{in}$.

For the average light curve and for the light curve in the high state $T_1$ was used in the form of the initial approximation and was accepted as the low limit for $T_{in}$. As the result of the minimization of the functional of residuals there were found values of six free parameters: $q$, $i$, $R_1$, $T_{in}$, $T_2$, $R_d$.

At the same time the value of $T_{in}$ significantly changed in the transition between low and high states, therefore the bolometric luminosity $L^c_{bol}$ of central parts of the disc (which provided the X-ray heating of the star and disc) also changed. This luminosity $L^c_{bol}$ can become comparable to the observed X-ray luminosity $L_x$ of the Sco~X-1 system (that can change from one epoch to another from $6\times 10^{36}$ erg s$^{-1}$ to $1.2\times 10^{38}$ erg s$^{-1}$).

\section{Results of modelling}

We performed modelling for the wide range of the parameter $T_1$, it was equal to $10^5$ K, $5\times 10^5$ K, $10^6$ K, $2\times 10^6$ K, $10^7$ K. Because of a significant lack of knowledge of specific values of the X-ray albedo of the star $\eta_s$ and disc $\eta_d$ along with the lack of precise knowledge of the disc's opening angle $\beta_d$ we considered at first the simplest case $\eta_s=\eta_d=0$, and for the opening angle we accepted the standard value from the theory of disc accretion \citep{shakura1973}: $\beta_d=3.2^{\circ}$.

\subsection{X-ray luminosities close to observed values}

\begin{figure}
\center
\includegraphics[width=\columnwidth]{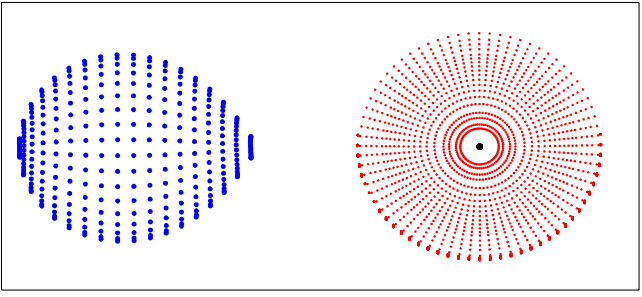}
\vspace{5pt} \caption{A computer model of the system calculated using optimal values of free parameters, the optical star fully filled its Roche lobe ($\mu=1$). Following values of parameters were used: $T_1=10^6$~K, $q=3.5$, $i=20^{\circ}$, $R_d=0.344a_0=0.56\xi$. The hot line and the hot spot were not shown, because their contributions in the total flux were insignificant.}\label{c-f-3}
\end{figure}

At first the interpretation of optical light curves of Sco X-1 was considered. Two values of $T_1$ were used ($5\times 10^5$ K and $10^6$ K), they corresponded (for $R_1$ values found below) to bolometric luminosities of the central source (in the disc) in the low state $L^c_{bol}=3.3\times 10^{36}$ erg s$^{-1}$ and $L^c_{bol}=2.6\times 10^{38}$ erg s$^{-1}$. These values were close to minimum and maximum values of the observed X-ray luminosity of Sco~X-1 ($L_x=6\times 10^{36}$ erg s$^{-1}$ and  $L_x=1.2\times 10^{38}$ erg s$^{-1}$ correspondingly). A computer model of the system was shown in Figure \ref{c-f-3}, it was computed using optimal values of free parameters.

\begin{figure}
\center
\includegraphics[width=\columnwidth]{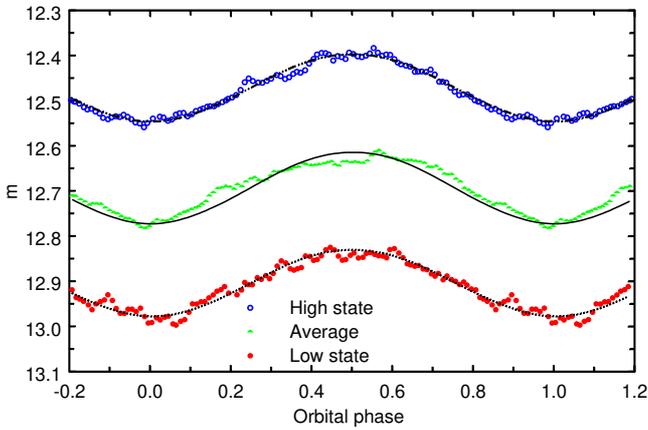}
\vspace{5pt} \caption{Observed light curves of Sco~X-1 with superimposed optimal theoretical light curves (synthesized using parameters from Table \ref{tab1}, $T_1=10^6$ K, $\mu=1$) in both states (low and high) and for the average curve.}\label{c-f-4}
\end{figure}

In Figure \ref{c-f-4} there were shown observed light curves of Sco~X-1 with superimposed optimal theoretical light curves in both states (low and high) and for the average curve. It can be seen that, despite of the fact that observations and theory were in a good agreement in general, in several parts there were discrepancies. This was due to complicated physical processes in the system that were not accounted for in our mathematical model. For normally distributed points in the light curve minimum residuals $\Delta_{\textrm{min}}$ should be distributed by the law $\chi^2_{N-M}$, where $N$ was the quantity of ``normal'' points within the light curve, $M$ was the quantity of free parameters used for minimization of residuals.

As was noted by \citet{hynes2016} the application of standard statistical criteria based on the $\chi^2$ statistics cannot be entirely justified, because the observed points in Kepler K2 mission potentially were distributed non-normally. Nevertheless, we formally used $\chi^2$ criterion for more clear visualization of results.

Calculations showed that minimum residuals $\Delta_{\textrm{min}}$ corresponded to $\chi^2=\frac{\Delta_{\textrm{min}}}{N-M}>>1$ (from 3 to 6). This fact indicated that our mathematical model was very simplified. Nevertheless, it was possible to choose optimal values of free parameters using minimum residuals $\Delta$ (if the sensitivity of the problem to changes of free parameters was enough). Namely in this way we understood all found values of parameters of the model.

\begin{figure}
\center
\includegraphics[width=\columnwidth]{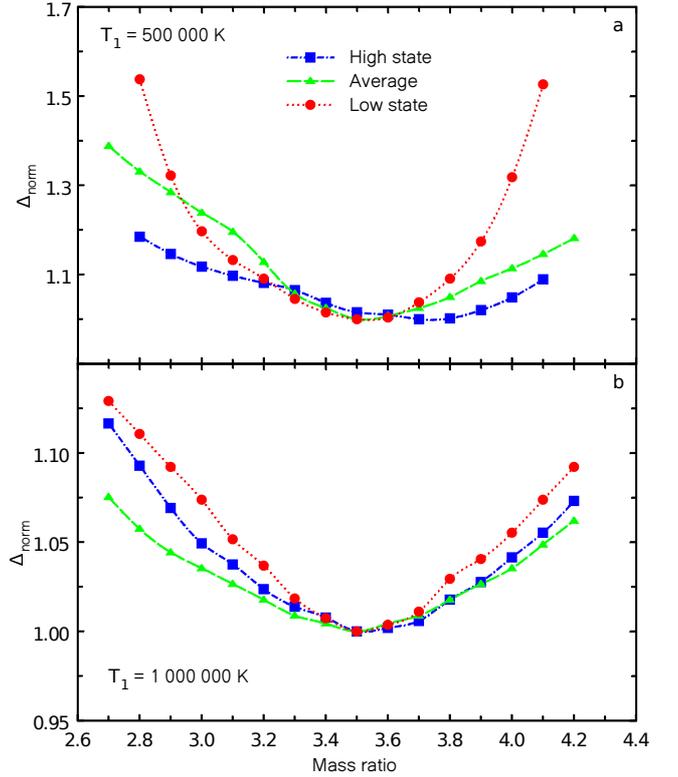}
\vspace{5pt} \caption{Dependencies of relative residuals $\Delta_{norm}=\chi^2/\chi^2_{\textrm{min}}$ minimized over all parameters except the mass ratio $q$ for $T_1=5\times 10^5$ K and $T_1=10^6$ K.}\label{c-f-5}
\end{figure}

In Figure \ref{c-f-5} we showed the dependence of residuals on the mass ratio $q$ minimized over all parameters (except $q$). Residuals $\Delta_{norm}$ were normalized using the minimum value $\Delta_{\textrm{min}}$ among values computed for different $q$. Results were shown for two different values of the $T_1$ parameter ($T_1=5\times 10^5$ K and $T_1=10^6$ K), in low and high states and for the average curve. It can be seen that all three curves demonstrated a significant dependence on the mass ratio. For $T_1=10^6$ K the value of $q$ was $q=3.5$, for $T_1=5\times 10^5$ K it was the same. As was noted above since our model was incompletely adequate and the distribution of observational points of light curves was not normal we showed only optimal values of parameters $q$ and $i$ obtained by the minimum of residuals and we did not show their errors. It was remarkable that the optimal value of $q$ did not depend (or depended weakly) on the value of $T_1$ (i.e., on the X-ray heating). It can be easily understood: the X-ray heating was mostly determined by the geometrical factor (by relative dimensions of the donor star that filled its Roche lobe, the size of the lobe depended on $q$).

\begin{figure}
\center
\includegraphics[width=\columnwidth]{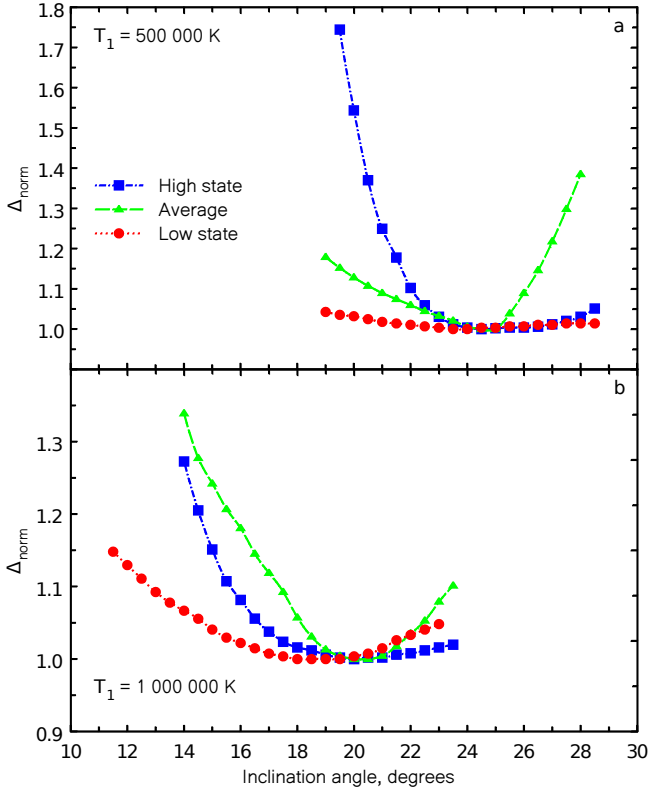}
\vspace{5pt} \caption{Dependencies of relative residuals $\Delta_{norm}=\chi^2/\chi^2_{\textrm{min}}$ minimized over all parameters except the orbital inclination $i$ for $T_1=5\times 10^5$ K and $T_1=10^6$ K.}\label{c-f-6}
\end{figure}

In Figure \ref{c-f-6} we showed the dependence of residuals $\Delta_{norm}$ on the orbital inclination $i$ (for $T_1=10^6$ K and $T_1=5\times 10^5$ K) minimized over all parameters (except $i$). Residuals also were shown for three variants of the light curve. All curves were similar and showed a significant dependence on $i$. For $T_1=10^6$ K the inclination was $i\approx 20^{\circ}$. For $T_1=5\times 10^5$ K it was $i\approx 25^{\circ}$. It can be seen that (in contradiction with the parameter $q$) the $i$ parameter significantly depended on $T_1$ (i.e., on the X-ray heating). It was naturally: if the value of X-ray heating was increasing the difference between average temperatures of heated and non-heated parts of the donor star (the value of $i$ was fixed) also was increasing, it led to the increasing of the ``reflection effect''. We did not give errors of $q$ and $i$, because our mathematical model was not completely adequate to observations.

\begin{table*}
\large
\centering
\caption{Optimal values of parameters of Sco X-1, obtained in modelling of optical light curves of Kepler K2 mission for low and high states and for the average curve, $T_1=10^6$ K. $T_2$ was the effective temperature of the donor star, $<T_{warm}>$ was the average temperature of this star on the heated part, $R_d$ was the radius of the disc in units of $\xi$ and $a_0$, $T_{in}$ was the average temperature of the disc's matter within the sphere with the $R_1$ radius, $T_{out}$ was the same on the outer border, $\alpha_g$ was the parameter that determined the temperature distribution along the disc's radius according to Equation \ref{eq-2}.}
\label{tab1}
\begin{threeparttable}
\begin{tabular}{@{}cccc@{}}
\hline
Parameters & Low state & Average & High state \\
\hline
$T_2$, K & $3 050$ & $3050$ & $3050$ \\
$<T_{warm}>$, K & 20~715 & 24~700 & 27~565 \\
$R_d$, $\xi$ & $0.535$ & $0.555$ & $0.606$ \\
$R_d$, $a_0$ & $0.332$ & $0.344$ & $0.376$ \\
$T_{in}$, K & $1\, 000\, 000$ & $1\, 204\, 310$ & $1\, 351\, 235$ \\
$T_{out}$ , K & $25\, 600$ & $28\, 815$ & $28\, 460$ \\
$\alpha_g$, fixed & 0.75 & 0.75 & 0.75 \\
$\chi^2$ & 292 & 329 & 503 \\
\end{tabular}
\begin{tablenotes}
\small
\item {\bf Note:} Parameters of synthetic light curves were obtained using following fixed values of parameters (they were calculated in previous stages of the work): the mass ratio was $q=M_x/M_v=3.5$, the orbital inclination was $i=20^{\circ}$, the average radius of the donor star was $\langle R_2\rangle$=0.286$a_0$, the disc's eccentricity was $e=0.01$, the argument of pericenter of the disc was $\alpha_e=110^{\circ}$, the distance between the inner Lagrange point $L_1$ and the center of masses of the neutron star was $\xi=0.6255 a_0$, the radius of the central sphere in the disc was $R_1=0.00314\xi = 0. 00196 a_0$, $T_1=10^6$ K was the temperature of the matter around $R_1$ distance in the low state, the thickness of the outer border of the disc was $\beta_d=3.2^{\circ}$, and it was assumed that the flux in ``arbitrary units'' $F_{12.7}=2.4558\times 10^{-8}$ corresponded to the stellar magnitude in white light $12.7^m$. 
\end{tablenotes}
\end{threeparttable}
\end{table*}

In Table \ref{tab1} we showed optimal values of Sco~X-1 parameters calculated for $T_1=10^6$ K. It can be seen that the temperature of the non-heated part of the donor star was $T_2=3050$ K, it corresponded to a spectral type M4-M5 \citep{habets1981}. The average radius of the optical star was close to  $0.286a_0$, where $a_0$ was the radius of the relative orbit, which can be estimated using Kepler's third law. Assuming the mass of the neutron star to be equal to its standard value $1.4 M_{\odot}$ the mass of the donor star was found to be $0.4 M_{\odot}$ for $q=3.5$. For the total mass of both components $1.8 M_{\odot}$ and for the orbital period $0.787$ d the radius of the relative orbit was $a_0=4.37R_{\odot}$ and the absolute average radius of the donor star $R_2=1.25 R_{\odot}=8.7\times 10^{10}$ cm. The bolometric luminosity of the donor star was $L_2=0.114 L_{\odot}$ for the temperature $T_2=3000$ K. So, the donor star in Sco X-1 possessed a significant excess of the radius and luminosity, i.e., the star noticeably moved forward in its nuclear evolution. Since the nuclear time scale for the $0.4 M_{\odot}$ main sequence star should be much longer than the current age of the Universe, it was necessary to suppose that the initial mass of the donor star was $>0.8 M_{\odot}$. The decrease of the donor star's mass could happen due to the strong stellar wind stimulated by the X-ray heating \citep{basko1973,iben1995}.
Another possibility of the donor star evolution can be associated with its deviation from thermal equilibrium due to relatively high mass loss rate. The characteristic time of mass loss by the star can become shorter than the time of thermal relaxation, as the result of this deviation the star should increase its radius possessing weakly evolved core (see, e.g., \citealp{knigge2006}). The accretion disc's radius in average was $R_d=(0.54-0.61)\xi=(0.33-0.38)a_0$ for $T_1=10^6$ K, where $\xi=0.626a_0$ was the distance between the disc's center and the inner Lagrange point L1, $a_0$ was the radius of the relative orbit. In the low state $R_d=0.54\xi$, in the high state $R_d=0.61\xi$, they coincide within errors. The temperature $T_{in}$ of the inner part of the accretion disc was $1.35\times 10^6$ K, $1.2\times 10^6$ K and $10^6$ K for the high state, for the average curve and for the low state correspondingly. The obtained value of the radius of the radiating central part of the disc (the central sphere in our model) was $R_1/a_0=0.00196$, i.e., $R_1= 6\times 10^8$ cm. The bolometric luminosity of the central part of the disc (it provided the X-ray heating of the star and disc) was $L^c_{bol}=8.4\times 10^{38}$ erg s$^{-1}$, $5.3\times 10^{38}$ erg s$^{-1}$ and $2.5\times 10^{38}$ erg s$^{-1}$, respectively. These values were somewhat greater than the upper observed limit of the Sco X-1 X-ray luminosity ($L_x=1.2\times 10^{38}$ erg s$^{-1}$) and the Eddington limit for the neutron star. Nevertheless, in methodological terms luminosity values clearly illustrated the cause of the transition from the low state to the high state: in the model of physical variability of the central X-ray source this happened due to the increase of the X-ray luminosity of the central part of the accretion disc by a factor of $\approx 3.36$. In our case the X-ray heating effect was observed in the optical range, i.e., in the Rayleigh---Jeans part of the spectrum. Therefore during the transition between low and high states the observed average optical brightness of the system should grow by the quantity that should be approximately proportional to the ratio of corresponding inner temperatures. It should be several tens of percents, and this estimate fitted observations. It was essential to note that circumstances of the donor star irradiation can change because of the change of the accretion disc's thickness or because of the variability of the anisotropy of the radiation of the central source.

\begin{table*}
\large
\centering
\caption{The same as Table \ref{tab1} for the temperature of the matter at the $R_1$ radius in the low state $T_1=5\times 10^5$ K and the orbital inclination $i=25^{\circ}$. The radius of the central sphere in the disc was $R_1=0.00217\xi=0.00136a_0$.}
\label{tab2}
\begin{threeparttable}
\begin{tabular}{@{}cccc@{}}
\hline
Parameters & Low state & Average & High state \\
\hline
$T_2$, K & $2940$ & $2940$ & $2820$ \\
$<T_{warm}>$, K & 9260 & 10~550 & 11~230 \\
$R_d$, $\xi$  & $0.636$ & $0.637$ & $0.685$ \\
$R_d$, $a_0$ & $0.394$ & $0.395$ & $0.424$ \\    
$T_{in}$, K & $500\, 000$ & $580\, 000$ & $623\, 880$ \\
$T_{out}$, K & $10\, 890$ & $11\, 500$ & $11\, 700$ \\
$\alpha_g$, fixed & $0.713$ & $0.710$ & $0.702$ \\
$\chi^2$ & 264 & 313 & 494 \\
\end{tabular}
\begin{tablenotes}
\small
\item {\bf Note:} The assumed stellar magnitude in the white light was $12.7^m$, the flux that corresponded to it was $F_{12.7}=7.283\times 10^{-9}$ in ``arbitrary units''.
\end{tablenotes}
\end{threeparttable}
\end{table*}

In Table \ref{tab2} we showed parameters of the system for $T_1 = 5\times 10^5$ K. The temperature of the non-heated part of the donor star was $2800-2900$ K. The average radius of the optical star was $\langle R_2\rangle=0.286a_0$. Since the radius of the optical star depended only on $q$, and $q=3.5$ was the same for $T_1=5\times 10^5$ K and $10^6$ K, the value of $\langle R_2\rangle$ was also the same. The accretion disc's radius was $0.63-0.68\xi=(0.39-0.42)a_0$. The temperature of the inner part of the disc $T_{in}$ was $5\times 10^5$ K, $5.8\times 10^5$ K and $6.2\times 10^5$ K for the low state, for the average curve and for the high state, respectively. The obtained radius of the radiating central part of the disc $R_1=0.00136a_0\approx 4.13\times 10^8$ cm. The bolometric luminosity of the disc's central part (that provided the X-ray heating of the star and disc) was $1.9\times 10^{37}$ erg s$^{-1}$, $1.4\times 10^{37}$ erg s$^{-1}$ and $8\times 10^{36}$ erg s$^{-1}$ for the high state, for the average curve and for the low state, respectively. These values were close to the lower limit of the observed Sco X-1 X-ray luminosity ($L_x=6\times 10^{36}$ erg s$^{-1}$). In the model of the variability of the temperature of the central source the increase of the luminosity of the central part of the accretion disc in the transition between low and high states from $8\times 10^{36}$ erg s$^{-1}$ to $1.9\times 10^{37}$ erg s$^{-1}$ (i.e. by a factor of 2.38) led to the corresponding change of the average brightness of the system.

\begin{figure}
\renewcommand{\thefigure}{\arabic{figure}a}
\includegraphics[width=\columnwidth]{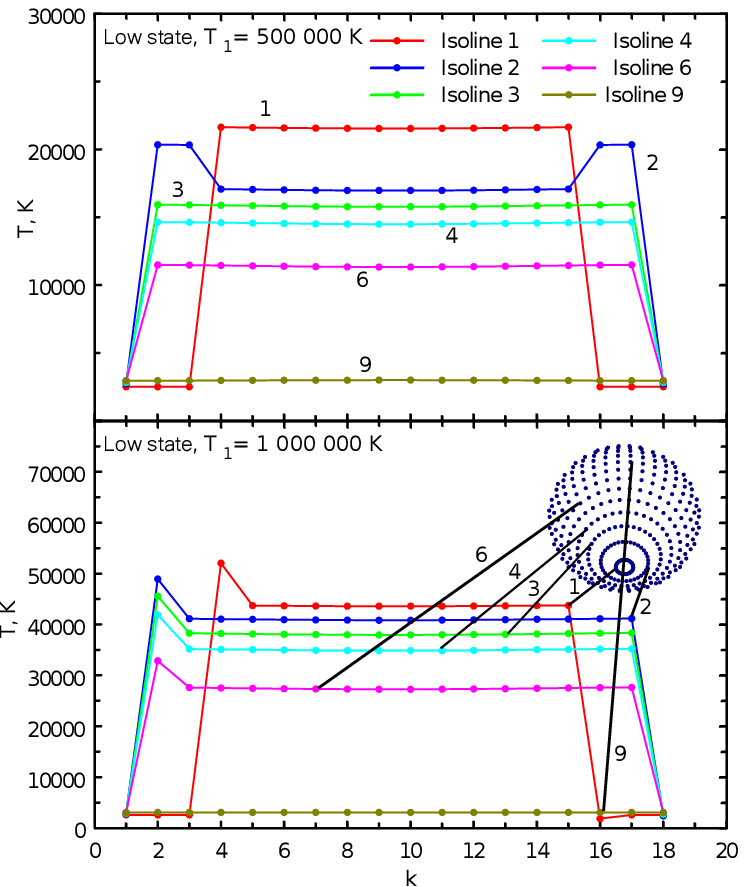}
\vspace{5pt} \caption{The distribution of the temperature on the heated (by X-rays) part of the donor star over the orbital plane for the low state, $T_1 = 5\times 10^5$ K and $T_1=10^6$ K. The region $\pm 5^{\circ}$ ($k=1-2$, $16-18$) around the L1 point ($j<1$) was shadowed by the accretion disc ($k$ was the index number of the area on the isoline $j$). This region corresponded to the angle $\varphi$ between the orbital plane and the radius-vector from the star's center of masses to the center of the elementary area from $5^{\circ}$ ($k=1$) to $95^{\circ}$ ($k=18$). $j$ showed the number of the isoline that was formed by elementary areas on the star. These areas with the same $j$ had the same angle $\theta$ between the axis that connected the centers of the components and the mentioned radius-vector. $j=1$ corresponded to $\theta=5^{\circ}$, $j=2$ was for $\theta=15^{\circ}$ and so on.}
\label{c-f-7-a}
\end{figure}

\begin{figure}
\renewcommand{\thefigure}{\arabic{figure}b}
\addtocounter{figure}{-1}
\includegraphics[width=\columnwidth]{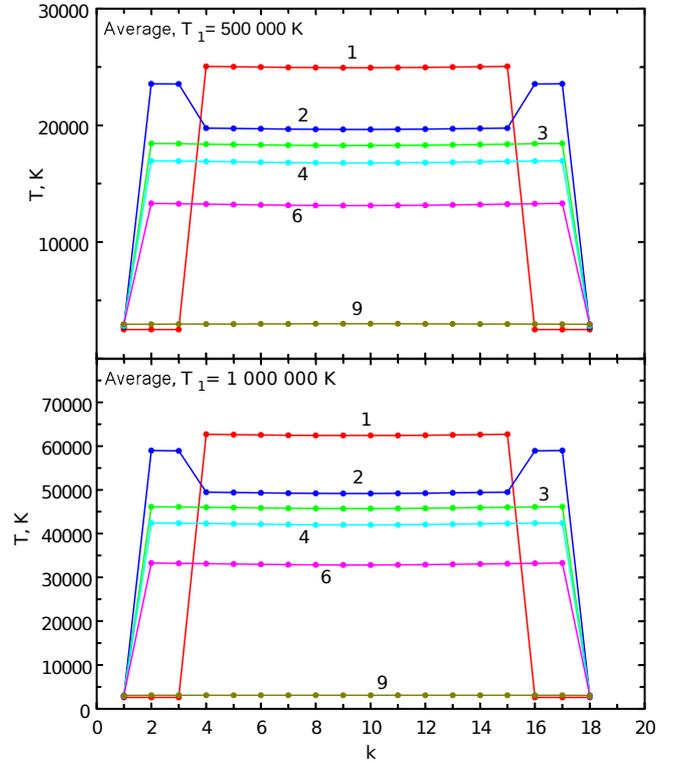}
\vspace{5pt} \caption{The same as Figure \ref{c-f-7-a} for the average curve.}
\label{c-f-7-b}
\end{figure}

\begin{figure}
\renewcommand{\thefigure}{\arabic{figure}c}
\addtocounter{figure}{-1}
\includegraphics[width=\columnwidth]{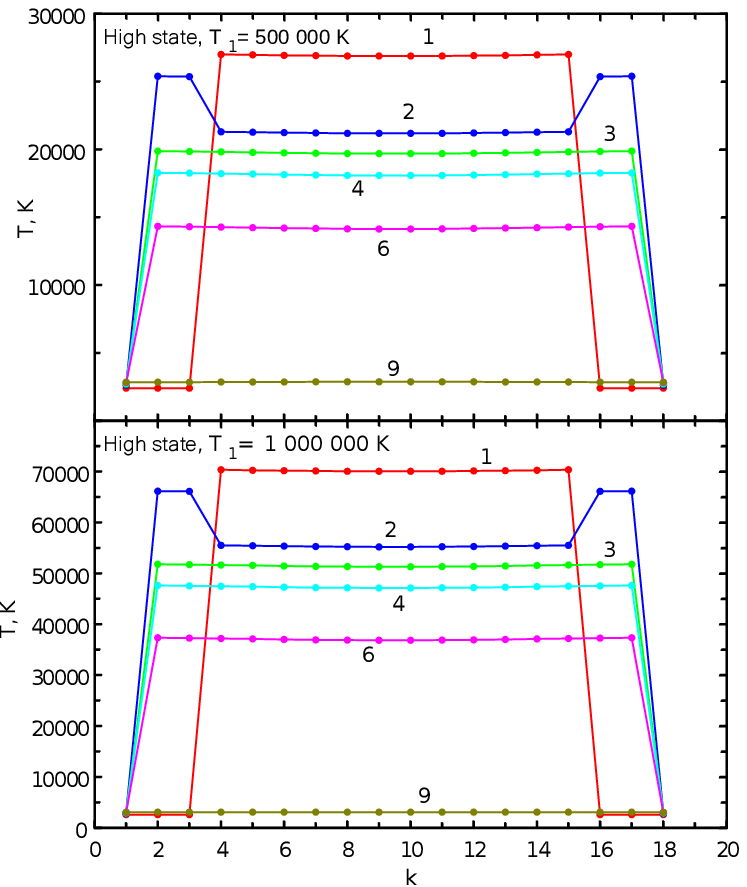}
\vspace{5pt} \caption{The same as Figure \ref{c-f-7-a} for the high state.}
\label{c-f-7-c}
\end{figure}

Figures \ref{c-f-7-a}, \ref{c-f-7-b} and \ref{c-f-7-c} demonstrated the temperature distribution on the heated (by X-rays) part of the optical star. Distributions of the temperature were shown for $T_1=5\times 10^5$ K and $T_1=10^6$ K, for the low and high states and for the average curve.

The zone around the Lagrange L1 point $\pm 5^{\circ}$ was shadowed by the accretion disc. This zone corresponded to the angle $\varphi$ between the orbital plane and the radius-vector from the star's center of masses to the center of the elementary area from $5^{\circ}$ to $95^{\circ}$. For the greater X-ray luminosity of the compact source ($T_1=10^6$ K) and for the high state the temperature decreased from $\approx 70\, 000$ K around the L1 point to $\approx 3000$ K in the vicinity of terminator (the terminator divided the heated and non-heated parts of the star's surface). The temperature about several tens thousands Kelvins was favourable for the formation of narrow emission lines NIII/CIII excited by Bowen mechanism in the starting point of the optical star's wind (the wind was induced by the X-ray heating, \citealp{basko1973}).

In case of the moderate X-ray heating ($T_1=5\times 10^5$ K) in the high state the temperature on the heated part of the optical star decreased from $\approx 27\, 000$ K around the L1 point to $\approx 3000$ K in the vicinity of terminator. As the result of the inverse problem solution (interpretations of light curves of Sco X-1) we computed light curves for the entire system and for its components: for the donor star, for the accretion disc, for the hot line and for the hot spot (contributions of the hot line and hot spot were negligible, therefore they were not shown separately throughout the manuscript). These curves can be expressed in units of the average total luminosity of the system or in arbitrary absolute energetic units. Arbitrary units can be converted to stellar magnitudes, in this case the flux of radiation for corresponding $T_1$ should be found using the average brightness of Sco X-1 ($12.7^m$, see above).

\begin{table*}
\large
\centering
\caption{Optical luminosities averaged for the orbital period (in the {\it C} band) of the disc and the stars of the system, expressed in units of the total average luminosity of Sco X-1 (Rel. flux) and in arbitrary absolute energetic units (a.u.), $T_1=10^6$ K, $i=20^{\circ}$.}
\label{tab3}
\begin{threeparttable}
\begin{tabular}{@{}ccc@{}}
\hline
Component & Abs. flux, $10^{-8}$ a.u. & Rel. flux \\
\hline
\multicolumn{3}{c}{Low state} \\
Optical star, $F_2$ & $0.32$ & $0.159$ \\
Disc's central sphere, $F_1$ & 0.0022 & 0.0011 \\
Disc, $F_d$ & $1.6320$ & $0.7997$ \\
Total flux, $F_{sys}$ & 2.04026 & 1 \\
\multicolumn{3}{c}{Average} \\
Optical star, $F_2$ & $0.42$ & $0.169$ \\
Disc's central sphere, $F_1$ & 0.0022 & 0.0009 \\
Disc, $F_d$ & $2.02100$ & $0.81580$ \\
Total flux, $F_{sys}$ & 2.47715 & 1 \\
\multicolumn{3}{c}{High state} \\
Optical star, $F_2$ & $0.49$ & $0.160$ \\
Disc's central sphere, $F_1$ & 0.0022 & 0.0007 \\
Disc, $F_d$ & $2.5070$ & $0.82550$ \\
Total flux, $F_{sys}$ & 3.03677 & 1 \\
\end{tabular}
\begin{tablenotes}
\small
\item {\bf Note:} Contributions of the hot line and hot spot to the flux were insignificant, therefore they were not shown.
\end{tablenotes}
\end{threeparttable}
\end{table*}

In Table \ref{tab3} we showed average luminosities of the disc and star during the orbital period in units of the total average luminosity of Sco X-1 and in arbitrary energetic units (a.e.u.), $T_1=10^6$ K. It can be seen that in both states and in the average curve the accretion disc's luminosity dominated. This luminosity decreased in the transition from the high state ($\approx 0.83$ a.e.u.) to the average curve ($\approx 0.82$ a.e.u.) and the low state ($\approx 0.80$ a.e.u.). In average the disc's luminosity was 4-5 times greater than the star's luminosity.

\begin{figure}
\includegraphics[width=\columnwidth]{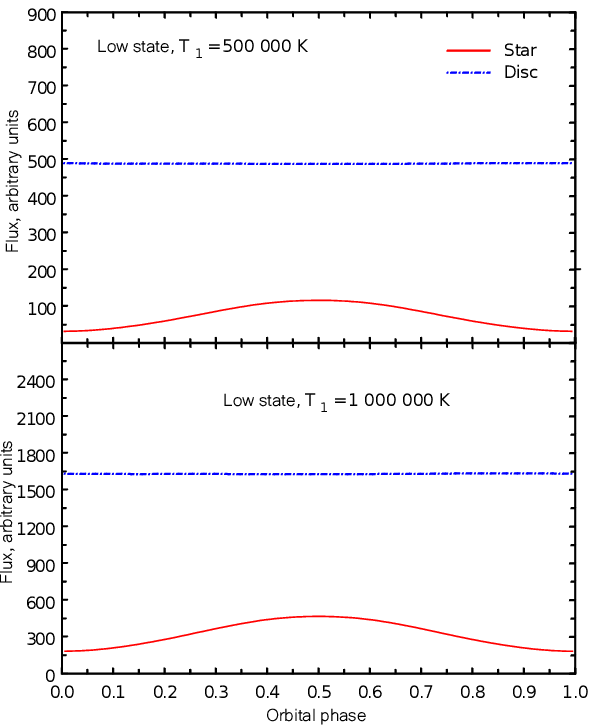}
\vspace{5pt} \caption{Light curves of the donor star and accretion disc calculated as the solution of the inverse problem, for the low state, $T_1$ parameter was $5\times 10^5$ K and $10^6$ K. The accretion disc dominated in of the total optical luminosity in all considered cases. }
\label{c-f-8-a}
\end{figure}

In Figure \ref{c-f-8-a} we showed theoretical light curves of the star and disc for the low state. It can be seen that the disc's luminosity did not change with the phase of the orbital period (contributions of luminosities of the hot spot and the hot line were negligible). Average theoretical light curves of the star and disc and these curves for the high state and average light curve were qualitatively similar to curves in Figure \ref{c-f-8-a}.

Luminosities of the disc and star monotonically decreased in the transition between low and high states. The light curve of the star itself was a single wave during the single orbital period (a ``reflection effect'', see, e.g., \citealp{cherepashchuk1972,lyutyi1973}), the amplitude and average brightness of it increased in the transition between low and high states.

The amplitude of observational light curves of Sco X-1 practically did not change (despite the fact that the average brightness of the system changed by $0.4^m$) due to the influence of the accretion disc's radiation and due to the value of the amplitude of the optical star's light curve and its average brightness.

\begin{table*}
\large
\centering
\caption{The same as Table \ref{tab3}, $T_1=5\times 10^5$ K, $i=25^{\circ}$.}
\label{tab4}
\begin{tabular}{@{}ccc@{}}
\hline
Component & Abs. flux, $10^{-8}$ a.u. & Rel. flux \\
\hline
\multicolumn{3}{c}{Low state} \\
Optical star, $F_2$ & $0.74$ & $0.122$ \\
Disc's central sphere, $F_1$ & 0.0050 & 0.0008 \\
Disc, $F_d$ & $4.885$ & $0.8086$ \\
Total flux, $F_{sys}$ & 6.04101 & 1 \\
\multicolumn{3}{c}{Average} \\
Optical star, $F_2$ & $0.98$ & $0.134$ \\
Disc's central sphere, $F_1$ & 0.0050 & 0.0007 \\
Disc, $F_d$ & $6.1820$ & $0.84070$ \\
Total flux, $F_{sys}$ & 7.35423 & 1 \\
\multicolumn{3}{c}{High state} \\
Optical star, $F_2$ & $1.13$ & $0.125$ \\
Disc's central sphere, $F_1$ & 0.0050 & 0.0006 \\
Disc, $F_d$ & $7.802$ & $0.86650$ \\
Total flux, $F_{sys}$ & 9.00446 & 1 \\
\end{tabular}
\end{table*}

In Table \ref{tab4} we showed average luminosities of the disc and star during the orbital period in units of the total average luminosity of Sco X-1 and in a.e.u., $T_1=5\times 10^5$ K. It can be seen that the ratio of luminosities of the star and disc were similar to the case of $T_1=10^6$ K.

\begin{figure*}
\includegraphics[width=\textwidth]{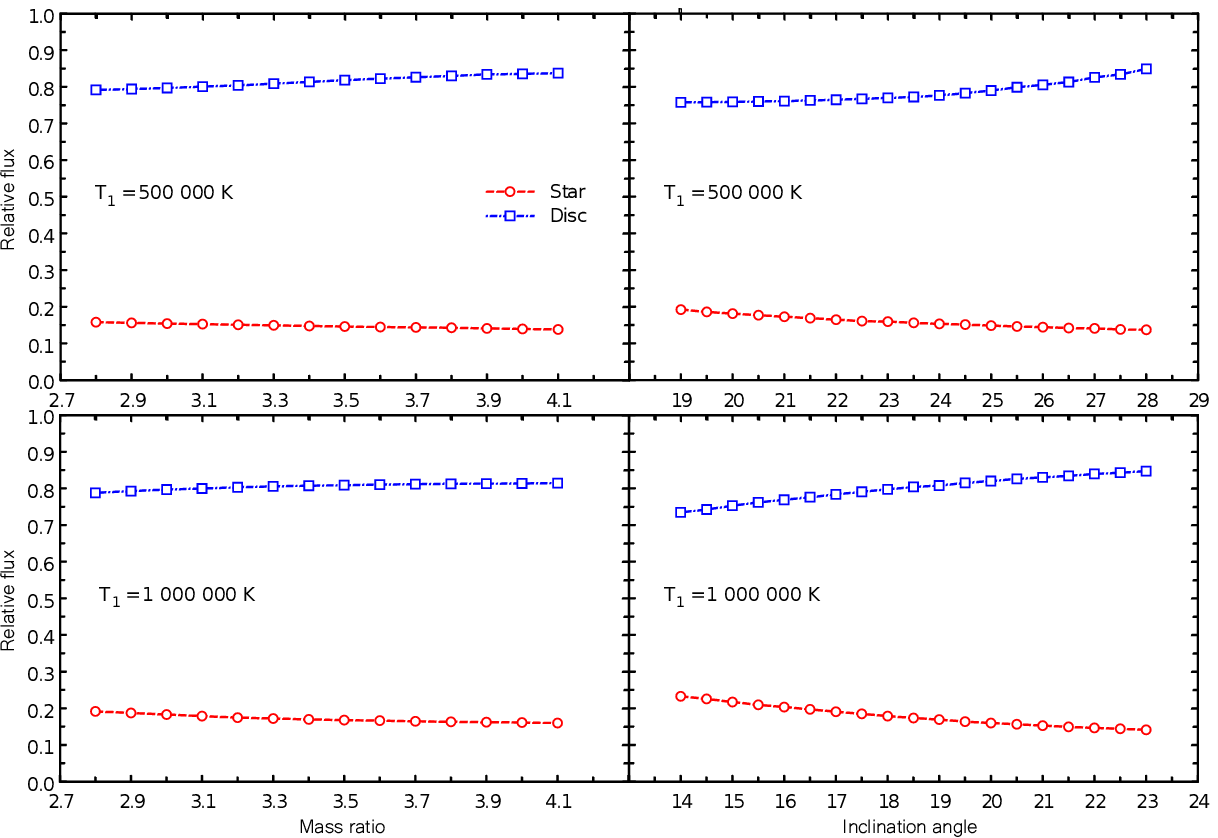}
\vspace{5pt} \caption{Dependencies of relative (averaged over the orbital cycle) luminosities of the accretion disc and donor star on parameters $q$ and $i$ for the average light curve, $T_1$ was equal to $5\times 10^5$ K and $10^6$ K.}
\label{c-f-9}
\end{figure*}

As followed from Figure \ref{c-f-9}, the ratio of the disc's and star's luminosities practically did not depend on values of free parameters, therefore the conclusion about the domination of the accretion disc's luminosity in the total optical luminosity of Sco X-1 was reliable.

\subsection{Wide range of X-ray luminosities}

The modelling of the Sco X-1 system in a wide range of luminosities of the X-ray source (much higher and much lower than observed values) was of great methodological interest. Additionally we interpreted light curves for $T_1=10^5$ K, $2\times 10^6$ K and $10^7$ K, system's luminosities in the low state were $1.9\times 10^{33}$ erg s$^{-1}$, $8\times 10^{39}$ erg s$^{-1}$ and $2\times 10^{43}$ erg s$^{-1}$. Values of X-ray luminosities significantly greater than $10^{38}$ erg s$^{-1}$ were physically unrealistic, because they were greater than the Eddington limit of the neutron star with mass $1.4 M_{\odot}$. Nevertheless, as noted above, we were used them from methodological reasons.

\begin{figure}
\includegraphics[width=\columnwidth]{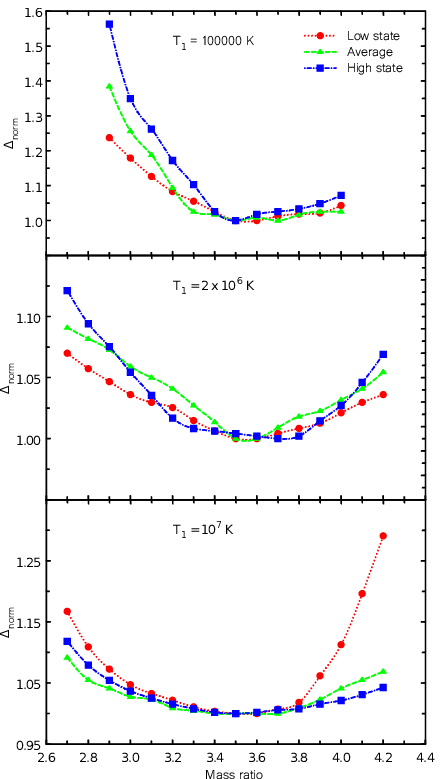}
\vspace{5pt} \caption{Dependencies of relative residuals $\Delta_{norm}=\chi^2/\chi^2_{\textrm{min}}$ minimized over all parameters except the mass ratio $q$ for $T_1=10^5$ K, $T_1=2\times 10^6$ K and $T_1=10^7$ K.}
\label{c-f-10}
\end{figure}

\begin{figure}
\includegraphics[width=\columnwidth]{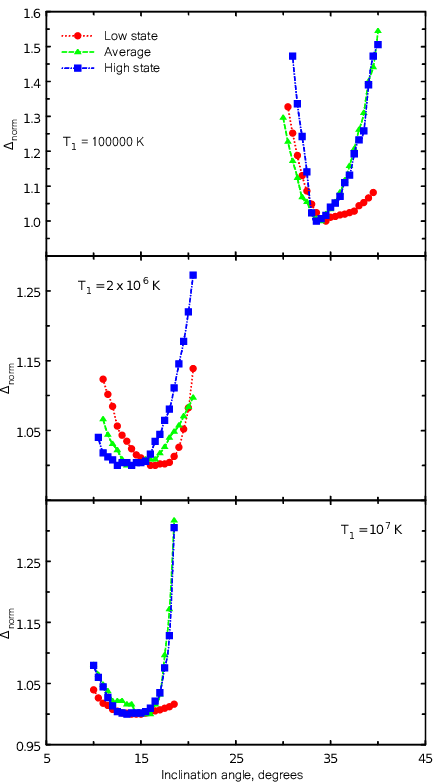}
\vspace{5pt} \caption{Dependencies of relative residuals $\Delta_{norm}=\chi^2/\chi^2_{\textrm{min}}$ minimized over all parameters except the orbital inclination angle $i$ for $T_1=10^5$ K, $T_1=2\times 10^6$ K and $T_1=10^7$ K.}
\label{c-f-11}
\end{figure}

Let us consider results of modelling for the low X-ray luminosity (the X-ray luminosity $L^c_{bol}$ of the sphere with $R_1$ radius, $L^c_{bol}=1.9\times 10^{33}$ erg s$^{-1}$ for $T_1=10^5$ K) that just slightly exceeded the bolometric luminosity of the donor star ($\approx 4.4\times 10^{32}$ erg s$^{-1}$). Even in such case the optimal mass ratio $q$ remained to be close to 3.5 (see Figure \ref{c-f-10}). However (see Figure \ref{c-f-11}), the orbital inclination angle $i=34^{\circ}$ that was significantly greater than $i=20^{\circ}$ (for $L^c_{bol}=(2.5-8.4)\times 10^{38}$ erg s$^{-1}$, $T_1=10^6$ K). The increase of the orbital inclination with the decrease of the X-ray heating can be easily understood. If the X-ray heating was low the ellipsoidal shape of the star also became important along with the ``reflection effect'' (which provided two waves in one orbital cycle). It led (in the same circumstances) to the decrease of the amplitude of the optical star's light curve, therefore it was necessary to increase $i$ to describe observed light curves.

Let us consider two variations of our modelling, they corresponded to very high X-ray luminosity: $(0.8-2.9)\times 10^{40}$ erg s$^{-1}$ ($T_1=2\times 10^6$ K) and $(2.0-8.7)\times 10^{43}$ erg s$^{-1}$ ($T_1=10^7$ K). As can be seen from Figure \ref{c-f-10}, the optimal mass ratio also remained the same for these very high X-ray luminosities: $q\approx 3.5$.

At the same time (as followed from Figure \ref{c-f-11}) the orbital inclination can be found using the minimum of residuals: $i\approx 15.5^{\circ}$ for $T_1=2\times 10^6$ K ($L^c_{bol}=8\times 10^{39}$ erg s$^{-1}$) and $i\approx 14.5^{\circ}$ for $T_1=10^7$ K ($L^c_{bol}=2\times 10^{43}$ erg s$^{-1}$). The temperature of the non-heated part of the optical star in all considered cases ($T_1=10^5$ K, $2\times 10^6$ K and $10^7$ K) was close to $2500-3000$ K. The temperature on the heated part of the star in the vicinity of the Lagrange point L1 was $\approx 1\, 200\, 000$ K, $\approx 140\, 000$ K and $\approx 3\, 200$ K for $T_1 = 10^7$ K, $2\times 10^6$ K and $10^5$ K correspondingly (for the average light curve).

\begin{figure}
\includegraphics[width=\columnwidth]{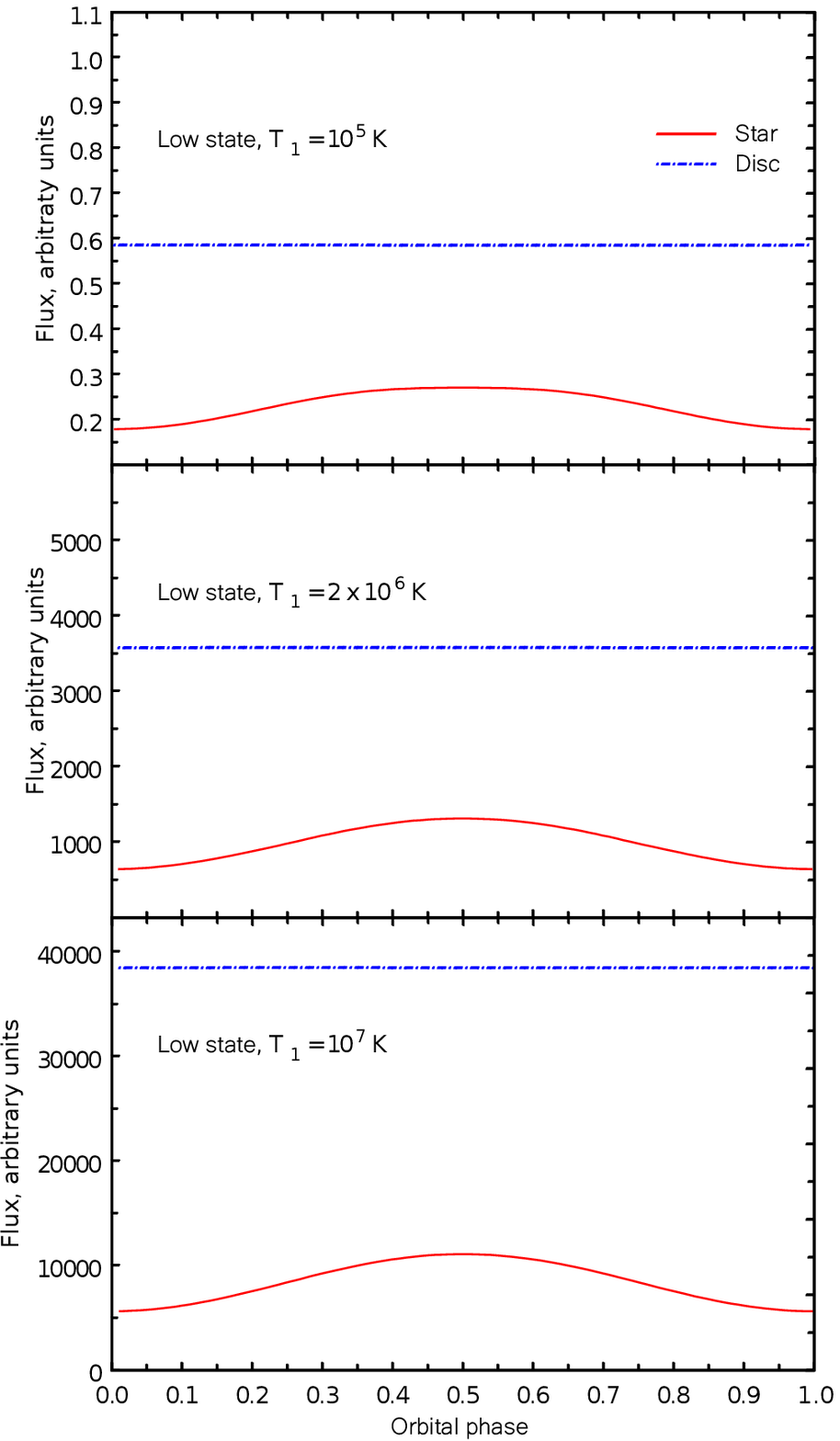}
\vspace{5pt} \caption{Light curves of the donor star and accretion disc computed as the inverse problem solution for the low state, $T_1=10^5$ K, $2\times 10^6$ K and $T_1=10^7$ K. Results for the high state and average curve were qualitatively similar to these curves.}
\label{c-f-12-a}
\end{figure}

As followed from Figure \ref{c-f-12-a} in cases of $T_1=10^5$ K, $2\times 10^6$ K and $10^7$ K all regularities (found earlier) remained: the contribution of the accretion disc dominated in the total optical radiation of Sco X-1. Thus, this domination effect did not depend on parameters $q$, $i$ and on the value of system's X-ray luminosity, because the solid angle of the accretion disc (that intercepted the X-ray radiation of the central source) was significantly greater that the solid angle of the optical star.

The most important results of our study were estimates of parameters $q$ and $i$. The value $q\simeq 3.5$ did not depend on the X-ray heating value and on the state of the system. At the same time the value of $i$ weakly depended on the system's state, but it strongly depended on the X-ray heating value.

\begin{table*}
\large
\centering
\caption{Parameters of Sco X-1 for different values of $T_1$ parameter (that characterized the X-ray heating in the case of $\eta_s=\eta_d=0$, $\beta_d=3.2^{\circ}$).}
\label{tab5}
\begin{threeparttable}
\begin{tabular}{@{}cccccccc@{}}
\hline
 & & & & & & & \\
$T_1$, & $R_1$, & $R_1$, & $L^c_{bol}=4\pi \sigma T^4_{in} R^2_1$, & $i$, & $q=\frac{M_x}{M_v}$ & $M_x^*$, & $\overline{K}_v^{**}$, \\
K & $a_0$ & cm & erg s$^{-1}$ & $^{\circ}$ & & $M_{\odot}$ & km s$^{-1}$ \\
\hline
$1\times 10^5$ & 0.00054 & $1.64\times 10^8$ & $1.9\times 10^{33}$ & 34 & 3.5 & 0.32 & 122 \\          
$5\times 10^5$ & 0.00136 & $4.13\times 10^8$ & $8.0\times 10^{36}$ & 25 & 3.5 & 0.75 & 92.6 \\
$1\times 10^6$ & 0.00196 & $5.96\times 10^8$ & $2.5\times 10^{38}$ & 20 & 3.5 & 1.42 & 74.9 \\
$2\times 10^6$ & 0.00269 & $8.18\times 10^8$ & $8.0\times 10^{39}$ & 15.5 & 3.5 & 2.97 & 58.5 \\
$1\times 10^7$ & 0.00575 & $1.75\times 10^9$ & $2.0\times 10^{43}$ & 14.5 & 3.5 & 3.61 & 54.8 \\
\end{tabular}
\begin{tablenotes}
\small
\item {\bf Note $^*$:} $M_x$ values were calculated using $K_v=74.9$ km s$^{-1}$ (i.e., $f_v(M)\approx 0.0343M_{\odot}$).
\item {\bf Note $^{**}$}: These values of $\overline{K}_v$ were required for definite values of $i$ under the assumption of the same value of the neutron star mass $M_x=1.4 M_{\odot}$. 
\end{tablenotes}
\end{threeparttable}
\end{table*}

\begin{figure}
\includegraphics[width=\columnwidth]{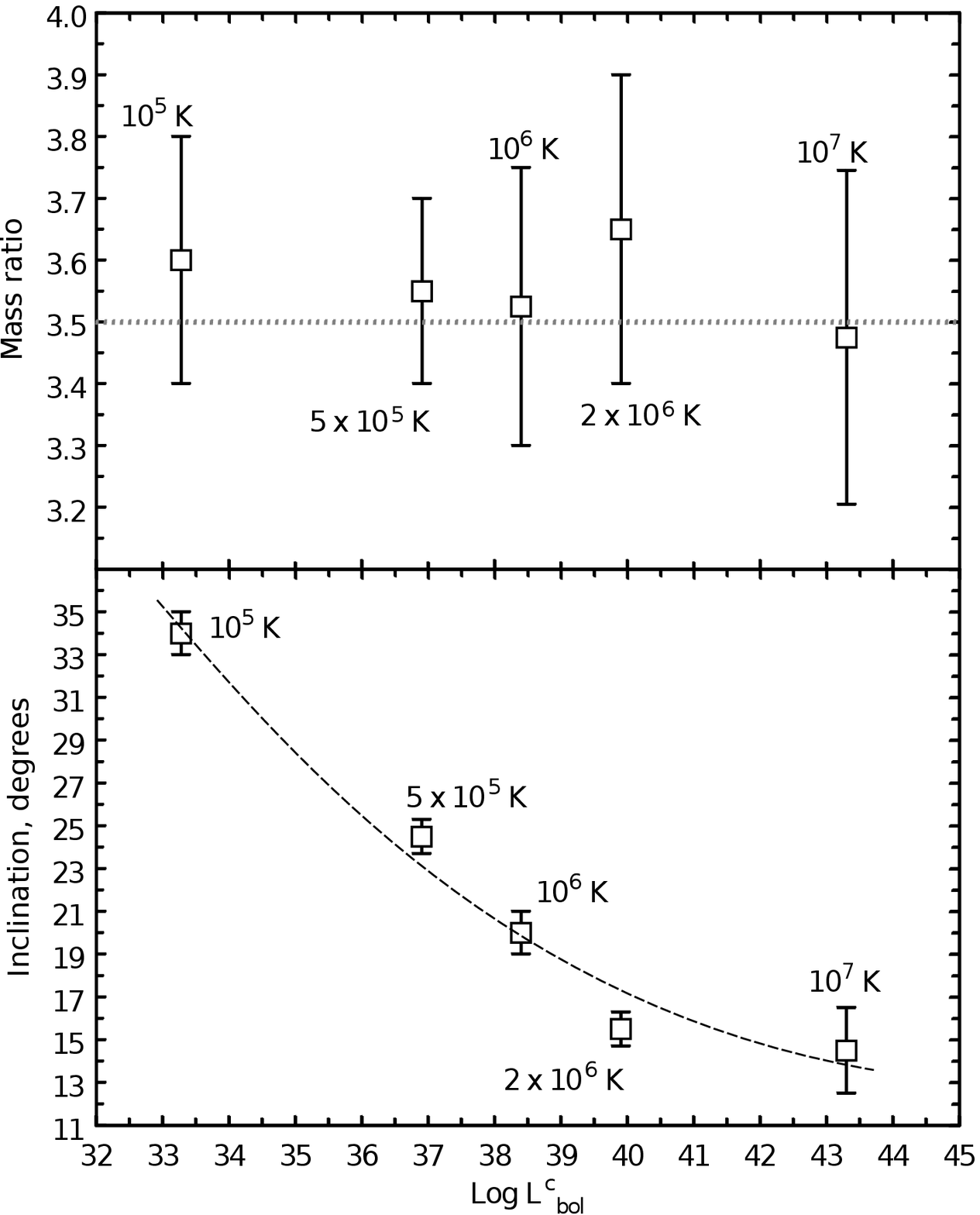}
\vspace{5pt} \caption{Dependence of $q$ and $i$ on the bolometric luminosity of the central part of the accretion disc in the low state (that characterized the X-ray heating). The corresponding value of $T_1$ was indicated. In the range $L^c_{bol}=3.3\times 10^{36}$---$2.6\times 10^{38}$ erg s$^{-1}$ ($\log L^c_{bol}=36.9$---$38.4$ erg s$^{-1}$, it was close to observed values of the Sco X-1 luminosity) the inclination angle of the orbit was $i=25-20^{\circ}$. The star fully filled its Roche lobe, $\eta_s=0$, $\eta_d=0$, $\beta_d=3.2^{\circ}$.}
\label{c-f-13}
\end{figure}

In Table \ref{tab5} and in Figure \ref{c-f-13} we showed the dependence of parameters $i$ and $q$ on the bolometric luminosity of the central part of the accretion disc in the low state (different values of $T_1$ were marked). The value $q\simeq 3.5$ was the same for all values of other parameters. The value of $i$ monotonically decreased from $i=34^{\circ}$ for $L^c_{bol}=1.9\times 10^{33}$ erg s$^{-1}$ to $i=14.5^{\circ}$ for $L^c_{bol}=2\times 10^{43}$ erg s$^{-1}$. In the range $L^c_{bol}=8.0\times 10^{36}-2.5\times 10^{38}$ erg s$^{-1}$ (that was close to the observed range of the X-ray luminosity of Sco X-1 $L_x=6\times 10^{36}-1.2\times 10^{38}$ erg s$^{-1}$) the value of $i$ was $25^{\circ}-20^{\circ}$.

\section{The influence of the thickness of the accretion disc and of its X-ray albedo on the light curve}

\subsection{The disc thickness}

Since the accretion disc lied in the orbital plane it shielded a part of X-ray flux leading to presence of a shadow on the surface of the optical star. We conducted the analysis (see above) of optical light curves of Sco X-1 using the standard theory of disc accretion \citep{shakura1973}. According its predictions for Sco X-1 parameters the opening angle of the standard accretion disc should be $\beta_d=\pm 3.2^{\circ}$. This value (thickness) was used in previous calculations above.

As was noted by \citet{dejong1996,dubus1999} the analysis of X-ray light curves of a number of low mass X-ray binaries led to the conclusion that the real value of the opening angle can be up to $12^{\circ}-14^{\circ}$ under the influence of the X-ray heating. Such high value of the disc's opening angle can be theoretically explained \citep{meyer1984}: the thickness of the outer part of the disc should significantly grow if the X-ray heating was strong. Moreover, the X-ray heating of the accretion disc by the central X-ray source can distort the disc's shape from symmetrical to curved \citep{pringle1996,maloney1996,maloney1997,dubus1999}.

\citet{pringle1996,maloney1996,maloney1997} showed that accretion discs irradiated by the central source (an accreting relativistic object) should be unstable against bending. It can be explained by the non-axisymmetric radiation pressure (in case of non-strictly flat discs) that can lead to disturbances in the disc and to the disc bending. Such curved accretion disc can precess under the influence of the force from the optical star with the period much longer than the orbital period as this took place in the Her X-1 X-ray binary.

The discovery of such long period variability in Sco X-1 would be a strong argument in favour of the curved disc model. In this model the shadow on the heated surface of the optical star can be caused by the shielding of the central X-ray source by the inner bent part of the accretion disc, but not by its outer part. Since such long period ``precessional'' variability'' was not found in Sco X-1, we considered symmetric accretion disc with different opening angles ($\beta_d=3.2^{\circ}-14^{\circ}$).

The investigation of the influence of the opening angle on results of our modelling was conducted for the average light curve of Sco X-1 using the value $T_1=5\times 10^5$ K and three fixed values of $\beta_d$ ($3.2^{\circ}$, $10^{\circ}$, $14^{\circ}$). In the first stage of calculations we assumed that the X-ray albedos of the optical star and accretion disc were $\eta_s=\eta_d=0$. It was made to separately study the influence of parameters $\beta_d$, $\eta_s$ and $\eta_d$ on results of the inverse problem solution. The method of the solution was the same.

\begin{figure}
\includegraphics[width=\columnwidth]{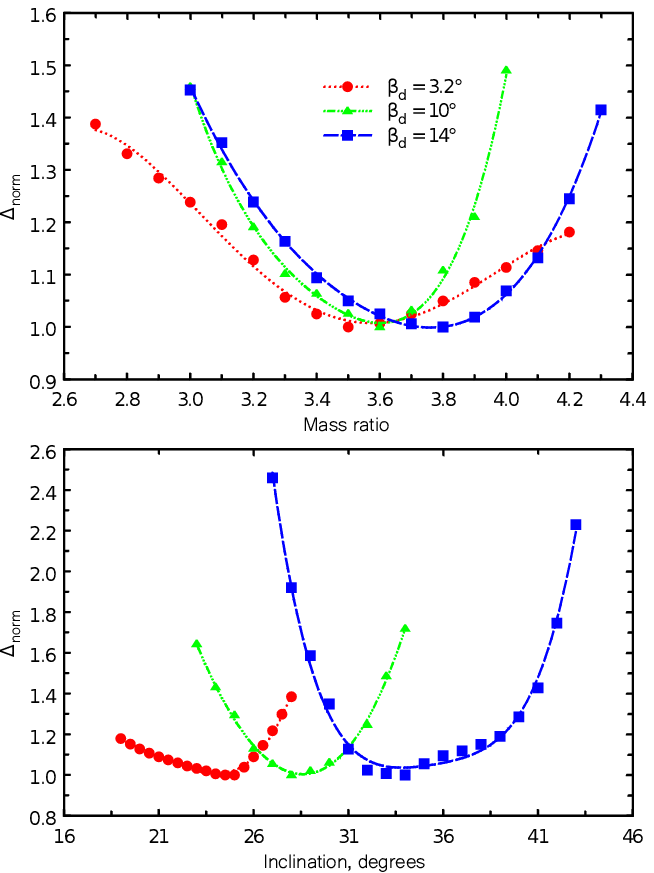}
\vspace{5pt} \caption{Dependencies of relative residuals minimized over all parameters except the mass ratio $q$ (upper panel) and inclination $i$ (lower panel) for the average light curve and different values of the disc's opening angle, $\eta_d=\eta_s=0$, $T_1=5\times 10^5$ K.}
\label{c-f-14}
\end{figure}

In Figure \ref{c-f-14} we showed dependencies of residuals between observed and theoretical light curves (normalized to minimal values) on parameters $q$ and $i$, the disc's opening angle was marked by quantities at corresponding curves. It can be seen that the influence of the opening angle on the optimal value of the mass ratio  $q=M_x/M_v$ was rather weak. The increase of $\beta_d$ from $3.2^{\circ}$ to $14^{\circ}$ led to the increase of $q$ from 3.5 to 3.8. At the same time the change of the opening angle led to a significant change of the orbital inclination $i$, for $\beta_d$ values $3.2^{\circ}$, $10^{\circ}$ and $14^{\circ}$ the values of $i$ were $25^{\circ}$, $28^{\circ}$ and $34^{\circ}$ correspondingly.

\subsection{The X-ray albedo}

\begin{figure}
\includegraphics[width=\columnwidth]{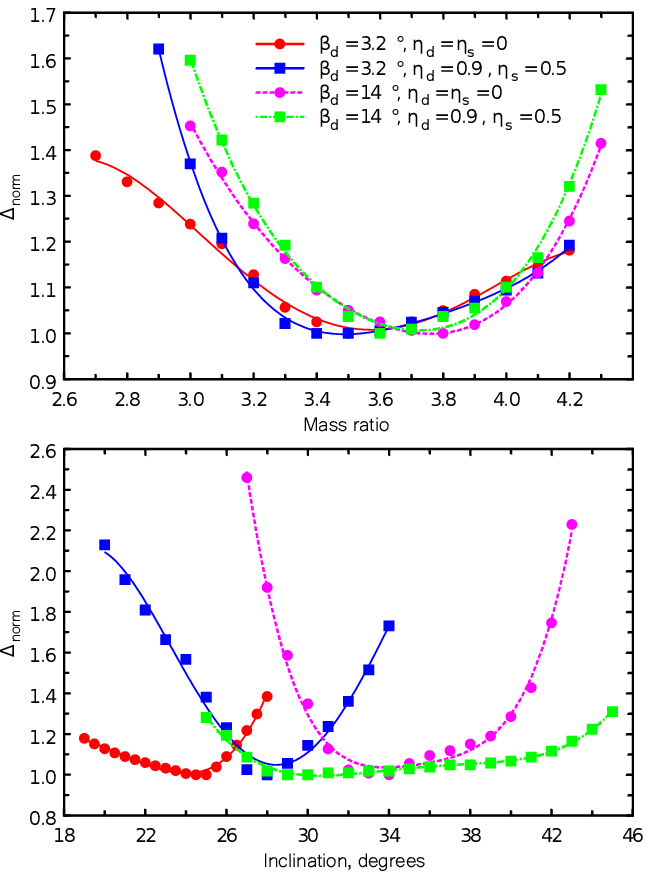}
\vspace{5pt} \caption{The same as Figure \ref{c-f-14} for following values of the disc's opening angle and X-ray albedo of the star and disc: $\beta_d=3.2^{\circ}$, $\beta_d=14^{\circ}$, $\eta_d=\eta_s=0$, $\eta_s=0.5$ and $\eta_d=0.9$.}
\label{c-f-15}
\end{figure}

\citet{dejong1996} estimated values of $\eta_s$ and $\eta_d$ from the comparison of a model of the X-ray heating of the accretion disc with observed X-ray luminosity of low mass X-ray binary systems. They showed that the X-ray albedo of the donor star can be accepted to be equal to 0.5, and for the accretion disc it was 0.9 (i.e., only about 10\% of the X-ray flux that fell from its central part to its outer part was thermally converted to the optical range). It can be supposed that other 90\% of falling flux was scattered by the outer layers of the disc without a significant change of the frequency or it was converted to the energy of convective motion in the disc. We analyzed the average light curve of Sco X-1 ($T_1=5\times 10^5$ K) using following values of parameters of the model: $\beta_d=14^{\circ}$, $\eta_s=0.5$, $\eta_d=0.9$. Results of this analysis (dependencies of residuals for parameters  $q$, $i$) were shown in Figure \ref{c-f-15}. It can be seen that the increase of the opening angle of the disc from $3.2^{\circ}$ to $14^{\circ}$ and the introduction the non-zero albedo ($\eta_s=0.5$, $\eta_d=0.9$) led to a weak change of the mass ratio ($q=3.6$) and to the value of the orbital inclination $i\approx 30^{\circ}$.

\begin{table*}
\large
\centering
\caption{Parameters of Sco X-1 for different values of $\beta_d$, $\eta_s$ and $\eta_d$, $T_1=500\,000$ K.}
\label{tab6}
\begin{tabular}{@{}cccc@{}}
\hline
 & \multicolumn{2}{c}{$\eta_s=\eta_d=0$} & $\eta_s=0.5$, $\eta_d=0.9$ \\
\hline
$q$ & 3.5 & 3.8 & 3.6 \\
$i$, $^{\circ}$ & 28 & 34 & 30 \\
$T_2$, K & 2950 & 2950 & 2950 \\
$<T_{warm}>$, K & 8170 & $10\,930$ & 8516 \\
$R_d$, $\xi$ & 0.400 & 0.420 & 0.362 \\
$R_d$, $a_0$ & 0.248 & 0.263 & 0.225 \\
$\beta_d$, $^{\circ}$ & 10 & 14 & 14 \\
$T_{in}$, K & $539\,300$ & $682\,550$ & $500\,930$ \\
$T_{out}$, K & $52\,600$ & $44\,860$ & $31\,700$ \\
$\alpha_g$, fixed & 0.75 & 0.726 & 0.70 \\
$R_1$, $\xi$ & 0.00200 & 0.00370 & 0.00362 \\
$R_1$, $a_0$ & 0.00126 & 0.00234 & 0.00227 \\
$\chi^2_{min}$ & 146 & 126 & 103 \\
\end{tabular}
\end{table*}

Table \ref{tab6} contained parameters of the Sco X-1 system that corresponded to optimal values of  $q$, $i$ for different values of the disc opening angle $\beta_d$ and albedos $\eta_s$, $\eta_d$. It can bee seen that for every fixed pair of $q$, $i$ values of parameters of the donor star and accretion disc were close to parameters found above for the average light curve for  $\beta_d=3.2^{\circ}$ and $\eta_s=\eta_d=0$ (see Table \ref{tab2}). The temperature of the non-disturbed star was $T_2\approx 3000$ K, the average temperature of the heated part of the star was  8000-11000 K, the radius of the accretion disc $R_d=(0.362-0.420)\xi$ was slightly less than in case of $\beta_d=3.2^{\circ}$, $\eta_s=\eta_d=0$ ($R_d=0.637\xi$), the temperature of its outer part was higher (32000-52600 K instead of 11500 K) because of the less radius and higher $\beta_d$.

The optical luminosity of the accretion disc dominated in the total optical luminosity of Sco~X-1. The contribution of the optical star was $\lesssim 20$\% as earlier (see Figure \ref{c-f-8-a}).

To find physical characteristics of the system we accepted following optimal values of parameters $q$, $i$ (taking into account Tables \ref{tab2} and \ref{tab6}): $q=3.6$ ($3.5-3.8$), $i=30^{\circ}$ ($25^{\circ}-34^{\circ}$), where in the brackets there were lower and upper limits of parameters defined mostly by the uncertainty of the physical model of the Sco X-1 system rather than by errors of observations.

\begin{figure}
\includegraphics[width=\columnwidth]{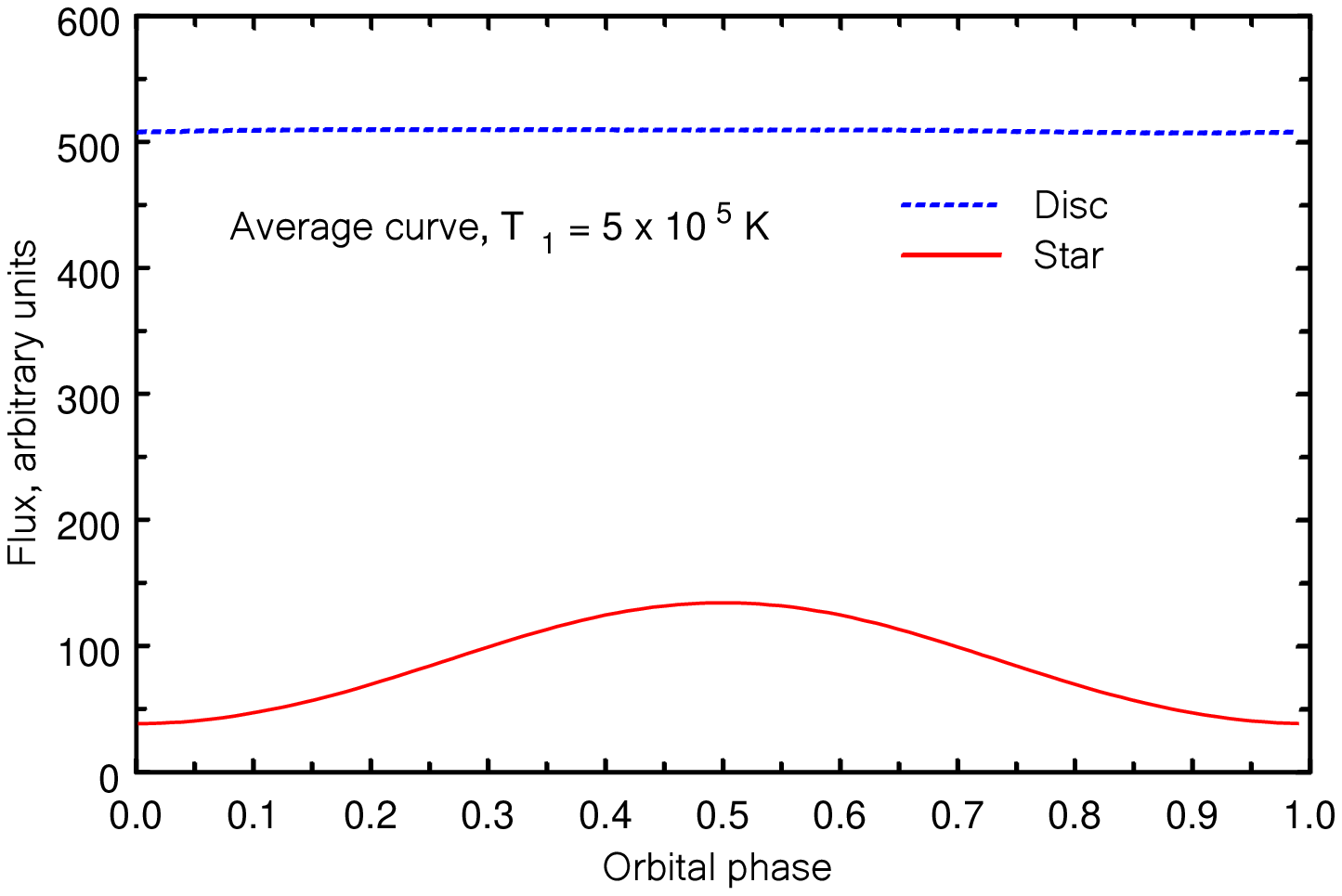}
\vspace{5pt} \caption{Light curves of the donor star and accretion disc computed as the inverse problem solution for the average light curve, $\eta_s=0.5$, $\eta_d=0.9$, $\beta_d=14^{\circ}$.}
\label{c-f-16}
\end{figure}

Figure \ref{c-f-16} showed theoretical light curves (in arbitrary units) of the donor star and accretion disc in case on the non-zero X-ray albedo and thick accretion disc ($\eta_s=0.5$, $\eta_d=0.9$, $\beta_d=14^{\circ}$).

\section{Discussion}

Let us to discuss results obtained in Section 4 ($\beta_d=3.2^{\circ}$, $\eta_s=\eta_d=0$), where the inverse problem was investigated in a wide range of X-ray luminosity $L_x$. In the case of low orbital inclination ($i=25^{\circ}-20^{\circ}$ that corresponded to the model in Section 4) the formation zone of Bowen lines should be close to the terminator of the heated star (where the temperature of the star's photosphere was low), whereas for high values of the orbital inclination this zone should be close to the Lagrange point (where the temperature of the star's photosphere was high), therefore we discussed here the case of low inclinations. The semi-amplitude of radial velocities of Sco X-1 obtained using narrow emission Bowen lines NIII/CIII was $K_v = 74.9$ km s$^{-1}$ \citep{galloway2014}. The corresponding mass function of the optical star was $0.0343 M_{\odot}$, it led to $M_x\geq 0.0569 M_{\odot}$ for $q=3.5$. Since Bowen lines were formed on the heated (by X-rays) part of the optical star, the value of $K_v=74.9$ km s$^{-1}$ was underestimated in comparison with the real $K_v$ that corresponded to the star's center of masses. So, the mass function $f_v(M)=0.0343 M_{\odot}$ was just a lower limit of the real mass function in the Sco X-1 system. In the model under discussion (with $\eta_s=\eta_d=0$, $\beta_d=3.2^{\circ}$) the knowledge of the values  of the inclination of the orbit $i=25^{\circ}-20^{\circ}$ and of the mass ratio $q=M_x/M_v=3.5$ allowed to estimate the lower limit of the mass of the relativistic object $M_x$ from the mass function $f_v(M)=0.034 M_{\odot}$: $M_x\geq 0.75 M_{\odot}$ for $i=25^{\circ}$ and $M_x\geq 1.42 M_{\odot}$ for $i=20^{\circ}$. It was interesting to note that for $i=20^{\circ}$ from our computed interval of $i$ the corresponding lower limit of the mass of the relativistic object was close to the average mass of the neutron star $1.4 M_{\odot}$. If we assumed this value of the relativistic object's mass we should assume that Bowen lines NIII/CIII were formed on the part of the photosphere of the optical star that was close to its center of masses. This was physically unrealistic, because (as was already mentioned) the temperature of the surface of the optical star around such region (close to terminator that separated the heated and non-heated parts of the star) was too low ($\approx 3000$ K) to excite ions of NIII/CIII. However, because of the high X-ray heating even in the vicinity of the terminator a zone with the inverse temperature distribution can form (like the solar chromosphere). In upper layers of such ``chromosphere'' physical conditions can become favourable for the formation of Bowen lines.

For the value $i=25^{\circ}$ from our computed interval of $i$ the corresponding mass of the relativistic object $M_x=0.75 M_{\odot}$ that was significantly lower than the neutron star's mass $1.4 M_{\odot}$. In order to extract the mass of the relativistic object from spectral data for $i=25^{\circ}$ that would be equal to the mass of the neutron star $1.4 M_{\odot}$ it was necessary to increase the observed semi-amplitude of radial velocities $K_v$ from 74.9 km s$^{-1}$ to 92.9 km s$^{-1}$, i.e. by a factor of 1.24. In this case the region of the Bowen lines formation did not coincide with the optical star's center of masses. The region was shifted from it in the direction of the relativistic object by $0.24a_v$, where $a_v$ was the radius of the absolute orbit the optical star. Since the absolute orbital radius $a_v$ and the distance between centers of masses of components $a_0$ were connected by the relation $a_v=a_0\dfrac{q}{1+q}\approx 0.78 a_0$ the shift of the region of Bowen lines origin from the center of masses of the optical star was $\approx 0.19a_0$. The distance between the center of star's masses and the L1 point was $0.37a_0$. As can be seen from Figures \ref{c-f-7-a}, \ref{c-f-7-b} and \ref{c-f-7-c} the location of the region of Bowen lines formation on the heated part of the star's surface in this case corresponded to $j=4-5$, where $j$ was the number of cross section of the star's body counted from the Lagrange point L1. For $T_1=10^6$ K the temperature at this place (see Figure \ref{c-f-7-c}) in the high state was $\approx 48\, 000$ K, for $T_1=5\times 10^5$ K it was $\approx 18\, 000$ K. The temperature in these regions on the heated part of the star significantly exceeded 10000 K allowing the formation of Bowen NIII/CIII lines. It was possible to move the region of Bowen lines formation even further from the optical star's center of masses (and, correspondingly, from the terminator) if we assumed the neutron star's mass in the Sco X-1 system exceeded the standard value $1.4 M_{\odot}$. Up to now there were found several binary systems with masses of pulsars (i.e., of neutron stars) were close to $2 M_{\odot}$ (see, e.g., \citealp{cherepashchuk2013}). If we assume that there was a massive neutron star in the Sco X-1 system ($\approx 2 M_{\odot}$) it was necessary for $i=25^{\circ}$ to increase the observed semi-amplitude of radial velocities from 74.9 km s$^{-1}$ to 104.9 km s$^{-1}$, i.e. by a factor of 1.4. In this case the region of the origin of narrow emission Bowen lines NIII/CIII was shifted from the center of the star's masses towards the relativistic object by $0.4a_v=0.31a_0$. It corresponded to $j=1-2$ (see Figures \ref{c-f-7-a}, \ref{c-f-7-b} and \ref{c-f-7-c}). The temperature here in the high state was $\approx 60\, 000$ K for $T_1=10^6$ K and $\approx 25\, 000$ K for $T_1=5\times 10^5$ K. High values of the temperature of the heated part of the optical star were favourable for the formation of Bowen lines NIII/CIII.

For the inclination $i=30^{\circ}$ ($25^{\circ}-34^{\circ}$) the zone of the formation of Bowen lines lied in the region with high temperatures of the photosphere of the heated optical star.

\section{Conclusions}

We interpreted optical orbital light curves of the Sco X-1 system using the model of the interacting binary system with the donor star filling its Roche lobe and the accretion disc around the relativistic object (neutron star).

We investigated our inverse problem in a wide range of parameters. The luminosity of the X-ray source $L_x$ was changed in a very wide range from $1.9\times 10^{33}$ erg s$^{-1}$ to $2.0\times 10^{43}$ erg s$^{-1}$ (see Table \ref{tab5}). The disc opening angle (that characterized the disc's thickness and defined the size of the shadow on the donor star) was changed from $\beta_d=3.2^{\circ}$ (the standard value in the disc accretion theory \citealp{shakura1973}) to $\beta_d=10^{\circ}$, $14^{\circ}$. Cases of the X-ray albedo of the disc $\eta_d=0$ and $\eta_d=0.9$ were considered. Main conclusions of our paper were following:

\begin{enumerate}

\item Calculations for the case $\beta_d=3.2^{\circ}$, $\eta_s=0$, $\eta_d=0$ and the range of values of the X-ray luminosity $L_x=1.9\times 10^{33}-2\times 10^{43}$ erg s$^{-1}$ showed that the mass ratio of components $q=M_x/M_v$ depended on the value of $L_x$ weakly, and it was $q\approx 3.5$. The orbital inclination $i$ significantly depended on the value of $L_x$ and it decreased from $i=34^{\circ}$ (for $L_x=1.9\times 10^{33}$ erg s$^{-1}$) to $i=14.5^{\circ}$ for ($L_x=2\times 10^{43}$ erg s$^{-1}$). For the physically realistic value $L_x=8\times 10^{36}$ erg s$^{-1}$ values of $q$ and $i$ were equal to 3.5 and $25^{\circ}$ correspondingly.

\item Calculation for the case $\eta_s=\eta_d=0$, $L_x=8\times 10^{36}$ for values of the accretion disc opening angle $\beta_d=10^{\circ}$, $14^{\circ}$ showed that the increase of this angle led to a significant increase of the orbital inclination $i$ and to a weak increase of the mass ratio $q$ (see Table \ref{tab6}). For the value $\beta_d=10^{\circ}$ corresponding values were $q=3.5$, $i=28^{\circ}$, for the value $\beta_d=14^{\circ}$ they were $q=3.8$ and $i=34^{\circ}$.

\item Calculation for $L_x=8\times 10^{36}$ erg s$^{-1}$ and for values $\beta_d=14^{\circ}$, $\eta_s=0.5$, $\eta_d=0.9$ gave following results: $q=3.6$, $i=30^{\circ}$. Finally, we chose following optimal values of parameters $q$, $i$: $q\simeq 3.6$ ($3.5-3.8$), $i=30^{\circ}$ ($25^{\circ}-34^{\circ}$), where in the brackets we showed lower and upper limits of $q$, $i$ that were mostly defined by the uncertainty of the physical model of the Sco X-1 system.

\item From the observational lower limit of the mass function of the optical star $f_v(M)=0.0343 M_{\odot}$ obtained from the semi-amplitude of the radial velocity for narrow Bowen lines $K_v=74.9$ km s$^{-1}$ for $q=3.6$, $i=30^{\circ}$ the estimates of masses of the relativistic object and optical star were $M_x>0.45 M_{\odot}$, $M_v>0.12 M_{\odot}$ correspondingly. To get the mass of the neutron star $M_x=1.4 M_{\odot}$ it was necessary to increase the semi-amplitude of the radial velocity of the optical star from $K_v=74.9$ km s$^{-1}$ to $K_v\approx 110$ km s$^{-1}$, i.e. by 1.47 times. It indicated that the zone of the formation of narrow emission Bowen lines in the Sco X-1 system was displaced from the center of mass of the optical star toward the Lagrange point $L_1$ by the value $0.44 a_v$, where $a_v$ was the radius of the absolute orbit of the optical star. In this region the temperature of the heated part of the optical star was significantly greater than 10~000 K, that was favourable for the formation of emission lines NIII/CIII. Besides, the temperature of the wind that moved from the star's heated part can be higher than on the photosphere level, and this fact also was favourable for the formation of Bowen lines.

\item For the value $q=3.6$ and for the neutron star's mass $1.4 M_{\odot}$ the possible mass of the optical star was $M_v\approx 0.4 M_{\odot}$. The temperature of the non-heated optical star $T_2\approx 3000$ K, it corresponded to the spectral type M4-M5V. The average radius of the optical star $R_2$ that filled its Roche lobe was $\approx 1.25 R_{\odot}$. The bolometric luminosity of the star was $L_{bol}=4\pi R^2_2 \sigma T^4_2=(2.1-4.6)\times 10^{32}$ erg s$^{-1}$. So, the optical star in the Sco X-1 system possessed significant excesses of radius and luminosity, i.e. it move in its evolutionary way and underwent a significant mass loss. Its initial mass should be greater than $0.8M_{\odot}$. Probably, the mass of the donor star was reduced due to the mass loss from the system in the stellar wind that was induced by the strong X-ray heating \citep{basko1973,iben1995}. Another possibility consisted in that the star was not in the thermal equilibrium and had an excess of the radius according to its mass due to the rapid mass loss.

\item The amplitude of the ``reflection effect'' for the donor star only (after the subtraction of the accretion disc's radiation) was maximal in the high state and decreased during the transition to the low state (and was lower for the average light curve). Since the isolated star cannot change its surface temperature distribution with time enough to explain the amplitude of the light curve we had strong reasons to claim that the main cause of the transition of Sco X-1 from the high state to the low state was the variability of the X-ray flux from the accreting neutron star. Under the assumption that the main cause of the change of the X-ray flux falling on the donor star was the change of the temperature in our modelling the X-ray flux dropped by a factor of 3.3 (for $T_1=10^6$ K) and by a factor of 2.4 (for $T_1=5\times 10^5$ K) during the transition between high and low states. It was essential to note that this result was related to the variability of Sco X-1 in long time scales. In short time scales the correlation between X-ray and optical fluxes in the system was more complicated \citep{hynes2016}: in the high state optical fluxes were correlated with X-ray fluxes, in the low state they were anti-correlated.

\item The optical luminosity of the accretion disc dominated in the total optical emission of Sco X-1 and was greater than the average luminosity of the donor star by a factor of $4$---$8$. It explained the invisibility of absorption lines in of the donor star in the system's spectrum and relative weakness of narrow Bowen emission lines \citep{galloway2014}. Optical luminosity of the accretion disc along with the amplitude of the ``reflection effect'' dropped during the transition from the high state to the low state, we explained this drop with the variability of the X-ray flux from the central source \citep{shakura1973}.

\item Our calculations showed that during the transition of Sco X-1 from the high state to the low state the total optical flux from the system dropped by $\approx 35$\%. The drop of the average flux from the donor star was $\approx 5$\%, and it was $\approx 30$\% for the accretion disc's flux. Contributions of the hot line and the hot spot were negligible ($\approx 1$\% of the total flux in the high state, $\approx 4$\% in the low state).

\item The ratio between optical fluxes of the accretion disc and of the donor star remained unchanged during the transition from the high state to the low state in a wide range of $q$ and $i$ parameters changes, i.e., our conclusion about the domination of the optical emission of the accretion disc in the total optical emission of Sco X-1 was reliable.

\end{enumerate}

The question about the nature of the bimodal behavior of the Sco X-1 system (the presence of low and high states in optical light curve differing by $0.4^m$) was a separate problem that was out the scope of this study. We made only several brief notes.

Slow changes of the average brightness combined with the orbital variability were observed in different types of binary systems. For example, in X-ray binaries with black holes in quiescence (when the X-ray luminosity was negligible) there were observed transitions from the passive state to the active state. During this process the average brightness of the system increased by several tenths of the stellar magnitude \citep{cantrell2008,cantrell2010,cherepashchuk2019a,cherepashchuk2019}. In the A0620-00 system the transition from the passive to the active state was accompanied with the start of the strong irregular variability of the brightness (flickering), in the system XTEJ1118+480 the growth of the average system's brightness was not accompanied with the flickering \citep{cherepashchuk2019a,cherepashchuk2019}. \citet{cherepashchuk2019a} suggested a hypothesis that the transition of the X-ray nova from the passive state to the active state and vice versa can be connected with movements of active regions of the donor star through the Lagrange point L1. Since the mass transfer rate through the L1 point strongly (as $(\Delta R/R)^3$) depended on the degree of the Roche lobe overflow $\Delta R$ \citep{paczynski1972} such movements can lead to significant changes of the rate of the matter consumption by the accretion disc and then to slow changes of the optical brightness of accretion structures.

In case of Sco X-1 such mechanism had low probability, because the influence of active regions on the donor star was suppressed by the strong X-ray heating.	It seemed that effects arose from the variability of the X-ray heating (they influenced on the rate of the mass transfer through L1 point) and effects arose from the interaction between winds from the star and disc (winds were induced by the strong X-ray heating, \citealp{basko1973}) became important. Also effects of eclipses of the central X-ray source by gas streams and structures in the accretion disc can be important, because they can lead to the decrease of the X-ray flux falling onto the optical star even if the luminosity of the X-ray source was constant. Therefore three dimensional gas dynamical calculations of such processes in binary X-ray systems were very actual.

\section*{Acknowledgements}

The work was supported by the Russian Science Foundation grant \mbox{17-12-01241} and by the Scientific and Educational School of M.~V.~Lomonosov Moscow State University ``Fundamental and applied space research'' (A.~M.~Cherepashchuk). The authors acknowledge support from M.~V.~Lomonosov Moscow State University Program of Development.

We are grateful to anonymous referee for valuable comments that helped to significantly improve the quality of the paper.

\section*{Data availability}

The data underlying this article will be shared on reasonable request to corresponding authors.

\bibliographystyle{mnras}
\bibliography{cherepashchuk-sco-x-1}

\bsp	
\label{lastpage}
\end{document}